\documentclass[dvipsnames]{vldb}
\usepackage{caption}
\usepackage{float}
\usepackage{courier}
\usepackage{url}
\usepackage{xspace}
\usepackage{listings}
\usepackage{graphicx}
\usepackage{mdframed}
\usepackage{graphics}
\usepackage{subcaption}
\usepackage{algorithm, algpseudocode}

\usepackage{microtype}

\usepackage{inconsolata}
\newenvironment{tightitemize}{%
\begin{list}{$\bullet$}%
{
\setlength{\itemsep}{0pt}%
\setlength{\topsep}{2pt}%
\setlength{\leftmargin}{0pt}%
\setlength{\parskip}{0pt}%
\setlength{\itemindent}{10pt}%
\setlength{\parsep}{2pt}
}}%
{\end{list}}

\newcounter{this-list}
\newenvironment{tightenumerate}{
\vspace{2pt}
\begin{list}{\arabic{this-list}.}{\usecounter{this-list}
                                 \setcounter{this-list}{0}
  \setlength{\itemsep}{0pt}%
  \setlength{\parsep}{0pt}%
  \setlength{\topsep}{0pt}%
  \setlength{\partopsep}{0pt}%
  \setlength{\leftmargin}{3.5ex}%
  \setlength{\labelwidth}{4.5ex}%
  \setlength{\labelsep}{1ex}%
}} {\end{list}\vspace{2pt}}

\newcommand{\hillview}{Hillview\xspace}
\newcommand{\code}[1]{\texttt{#1}}
\newcommand{\overlook}{Overlook\xspace}

\newcommand{\eps}{\varepsilon}

\newcommand{\DP}{differential privacy}
\newcommand{\A}{\mathcal A}

\newcommand{\K}{\mathcal K}
\newcommand{\X}{\mathcal X}
\newcommand{\Q}{\mathcal Q}
\newcommand{\Y}{\mathcal Y}
\newcommand{\M}{\mathcal M}
\newcommand{\Z}{\mathcal Z}
\def\H{\mathcal H}
\newcommand{\rgta}{\ensuremath{\rightarrow}}
\newcommand{\remove}[1]{}

\usepackage{mathtools}

\newcommand{\cf}{\emph{cf.}\xspace}
\DeclareMathOperator{\sgn}{sgn}

\newcommand{\Lap}{\ensuremath{\mathrm{Lap}}}

\newcommand{\hh}{hierarchical histogram\xspace}

\renewcommand{\paragraph}[1]{{\vspace{.45em}\noindent\textbf{#1.} }}

\title{\overlook: Differentially Private Exploratory Visualization for Big Data}
\vldbTitle{\overlook: Differentially Private Exploratory Visualization for Big Data}
\vldbAuthors{Pratiksha Thaker, Mihai Budiu, Parikshit Gopalan, Udi
Wieder, and Matei Zaharia}
\vldbDOI{}
\vldbVolume{12}
\vldbNumber{11}
\vldbYear{2020}

\numberofauthors{5}
\author{
\alignauthor
  Pratiksha Thaker \\
  \affaddr{Stanford University}
  \email{prthaker@stanford.edu}
  \alignauthor Mihai Budiu \\
  \affaddr{VMware Research}
  \email{mbudiu@vmware.com} 
  \alignauthor
  Parikshit Gopalan \\
  \affaddr{VMware Research}
  \email{pgopalan@vmware.com}
\and
  \alignauthor Udi Wieder \\
  \affaddr{VMware Research}
  \email{uwieder@vmware.com}
  \alignauthor
  Matei Zaharia \\
  \affaddr{Stanford University}
  \email{matei@cs.stanford.edu}
}

\begin{document}

\maketitle

\begin{abstract}
Data exploration systems that provide differential privacy must manage a privacy budget that measures the amount of privacy lost across multiple queries. One effective strategy to manage the privacy budget is to compute a one-time private synopsis of the data, to which users can make an unlimited number of queries. However, existing systems using synopses are built for offline use cases, where a set of queries is known ahead of time and the system carefully optimizes a synopsis for it. The synopses that these systems build are costly to compute and may also be costly to store.

We introduce Overlook, a system that enables private data exploration at interactive latencies for both data analysts and data curators. The key idea in Overlook is a \emph{virtual synopsis} that can be evaluated \emph{incrementally}, without extra space storage or expensive precomputation. Overlook simply executes queries using an existing engine, such as a SQL DBMS, and adds noise to their results. Because Overlook's synopses do not require costly precomputation or storage, data curators can also use Overlook to explore the impact of privacy parameters interactively. Overlook offers a rich visual query interface based on the open source Hillview system. Overlook achieves accuracy comparable to existing synopsis-based systems, while offering  better performance and removing the need for extra storage.

\end{abstract}

\section{Introduction}

Privacy has become a key issue for all organizations that collect personal data, from companies to government entities~\cite{us-census-privacy, apple-privacy, google-privacy}.
After organizations collect a dataset, they would like to make it available to internal data analysis teams, or even expose it to external researchers~\cite{us-census-privacy}, without leaking a significant amount of information about any individual in the dataset.
To be broadly useful, a private data analysis system should support ad-hoc, exploratory queries through familiar interfaces while making it easy for the data curator (the administrator configuring the system) to control the amount of information leaked.

The most widely used framework for reasoning about privacy is differential privacy (DP)~\cite{dp, dp-survey}.
Differential privacy quantifies the privacy cost of a statistical analysis through a \emph{privacy budget};
a smaller privacy budget implies more error in the query results but more privacy for the individuals in the dataset.

Although many research systems provide differential privacy~\cite{pinq, Johnson18, psi, privatesql, dawa, airavat}, these current systems are challenging for organizations to configure and use, especially for ad-hoc exploratory analysis.
At a high level, current DP systems fall into two categories:
\begin{tightenumerate}
  \item \textbf{Systems with per-query budgeting:} Systems such as PINQ~\cite{pinq} and FLEX~\cite{Johnson18} ask users to select a privacy budget, $\eps$, for each query they execute.
  The total privacy leakage of the system is then bounded by the sum of these $\eps$ values.
  These systems are complex for both users and data curators to use.
  Users typically have a limited total privacy budget available, $\eps_\mathrm{total}$, and need to decide how to divide it between the queries they submit; when they run out of budget, they can no longer make queries.
  In addition, two users that collaborate can reveal information proportional to the sum of their budgets, so data curators must carefully limit which users can access the system.
  Such systems would be unsuitable for exposing a research dataset to the public, for instance~\cite{us-census-privacy}.

  \item \textbf{Synopsis-based systems:} Systems such as PrivateSQL \cite{privatesql} generate a \emph{synopsis} data structure that can answer a specific class of queries after taking in a dataset, a description of the query class, and a total privacy budget $\eps$.
  Users can then query the synopsis arbitrarily many times without revealing additional information beyond the $\eps$ budget.
  These systems are more suitable for exploratory analysis and for public access, but unfortunately, they are also challenging to use.
  Constructing the synopsis requires solving an expensive optimization problem to minimize the error it will produce for a specific query workload, which can take hours even for a modest dataset, and the synopsis can consume a large amount of space, on par with the original data, making it costly for large datasets.
\end{tightenumerate}

In this paper, we present Overlook, a system that makes synopsis-based
differential privacy practical for one of the most common types of
data analysis: visual exploratory analysis of immutable datasets.
Visual query interfaces, such as Tableau~\cite{tableau}, are one of the most common ways for organizations to expose data internally, and produce a class of queries that are a good fit for synopsis data structures (mostly counting queries).
In Overlook, we seek to make private visual queries accessible to both data users and data curators, by designing a system that lets curators tune a synopsis \emph{interactively} to set privacy parameters, and lets users query data interactively at a similar cost to their existing data analysis infrastructure.
Overlook runs as an interposition layer in front of existing analytical engines, such as any SQL RDBMS, enabling organizations to benefit from the scalability and optimizations of these existing engines and to offer private visual query interfaces over existing datasets.

The key idea in Overlook is a \emph{virtual synopsis} data structure that represents the noise that would be added by a classical synopsis algorithm in a highly compressed format using a pseudo-random function (PRF).
For any counting query (e.g., counting the users in a dataset by country), Overlook can use the virtual synopsis to compute just the noise that should be added to each tuple in the query result.
Overlook simply adds this noise to the results from an existing query engine. 
Thanks to this design, users can run queries at a similar speed to their existing query engine, with extra computation that is only proportional to the number of output tuples (e.g., number of bins in a histogram).
Likewise, data curators can use Overlook to explore parameters of the virtual synopsis \emph{interactively}, e.g., change the total privacy budget $\eps$ and see its results on several queries.
Overlook's synopses are based on the \hh mechanism~\cite{Hay10, binary-mechanism}, a synopsis design that supports multidimensional queries, and can be tuned by curators to provide different noise levels for different dimensions in the data.

Overlook also offers a rich privacy-aware visual query interface built on virtual synopses, based on the open source Hillview system~\cite{hillview}.
In particular, we extend most of the built-in visualizations in Hillview, such as histograms and heatmaps, to display information about the noise introduced by DP, and to automatically coarsen the visualization when the noise exceeds the discernible signal.
These automatic adjustments all query the virtual synopsis, so they do not cause any additional privacy leakage.
Data visualization has some unique features that make
it a good candidate for a synopsis based approach to DP:  most
queries could be expressed as (combinations of) count queries, for which there are good
synopses. Secondly, the visualization itself introduces errors via the
quantization to the pixels in the screen. We found that this inherent
approximation often masks the error introduced by DP. Finally,
typically data visualization interfaces  already incorporate methods
for presenting errors and approximations to the user (e.g., via
confidence intervals and error bars). These tools help the user
understand the results the \DP~mechanism produces.

Unlike prior systems, Overlook provides an interface for both the data \emph{curator}
as well as the data \emph{analyst}.
The data curator, who manages and has access to the raw data,
must make decisions about the privacy parameters used to build the synopsis.
\overlook's \emph{curator UI} lets the curator quickly see the impact of adjusting various privacy parameters, such as bins for categorical features and privacy budgets along different data dimensions, on multiple visual queries.
To our knowledge, Overlook is the first system to provide interactive feedback for tuning a DP synopsis.

We implement Overlook using the Hillview UI, and develop backends to let it run either over a SQL DBMS or over Hillview's built-in distributed execution engine~\cite{hillview}, a high performance in-memory query engine that supports approximate query processing for common visualization queries.
In both cases, Overlook benefits directly from the optimizations in the underlying engine.

We evaluate Overlook against the algorithms for computing synopses in DAWA~\cite{dawa} and PrivateSQL~\cite{privatesql}, and other algorithms in DPBench~\cite{dpbench}.
Although many of these algorithms build workload-aware synopses, which are optimized for a specific set of queries~\cite{matrix-mechanism}, we find that Overlook's workload-agnostic virtual synopses offer similar levels of error in query results when given the set of visualization queries as input.
The key intuition is that in an exploratory analytics setting with many possible queries, a synopsis that ``balances'' the noise over the possible queries will perform well, and it is not useful to solve a complex optimization problem~\cite{QardajiYL13, qardaji2013differentially, Hay10}.
Moreover, Overlook's virtual synopses require no precomputation and minimal storage over the
underlying histogram.
We show that Overlook achieves the same scaling properties as the underlying database
while requiring no more than 2.5$\times$ the time required to compute equivalent non-private queries.
Overlook requires only a 32-byte key for its synopsis,
compared to existing synopses that require kilobytes of space to store
and potentially gigabytes of memory to compute.

To summarize, our contributions are:
\begin{tightenumerate}
  \item We present Overlook, the first differentially private visual analytics system that provides interactive configuration for data curators and simultaneously supports arbitrary, interactive ad-hoc queries for users by leveraging synopses.
  \item We introduce virtual synopses, a data structure to represent DP synopses for multidimensional counting queries that can be used incrementally to add noise to a specific query instead of requiring costly precomputation and storage.
  \item We develop a privacy-aware interactive visualization UI for both data users and data curators, including a novel curation UI that lets curators see the effect of configuring virtual synopsis parameters interactively.
\end{tightenumerate}

Overlook is open source at \url{http://github.com/vmware/hillview}.

\section{System overview}

\begin{figure}[t!]
  \begin{center}
    \includegraphics[width=.95\columnwidth]{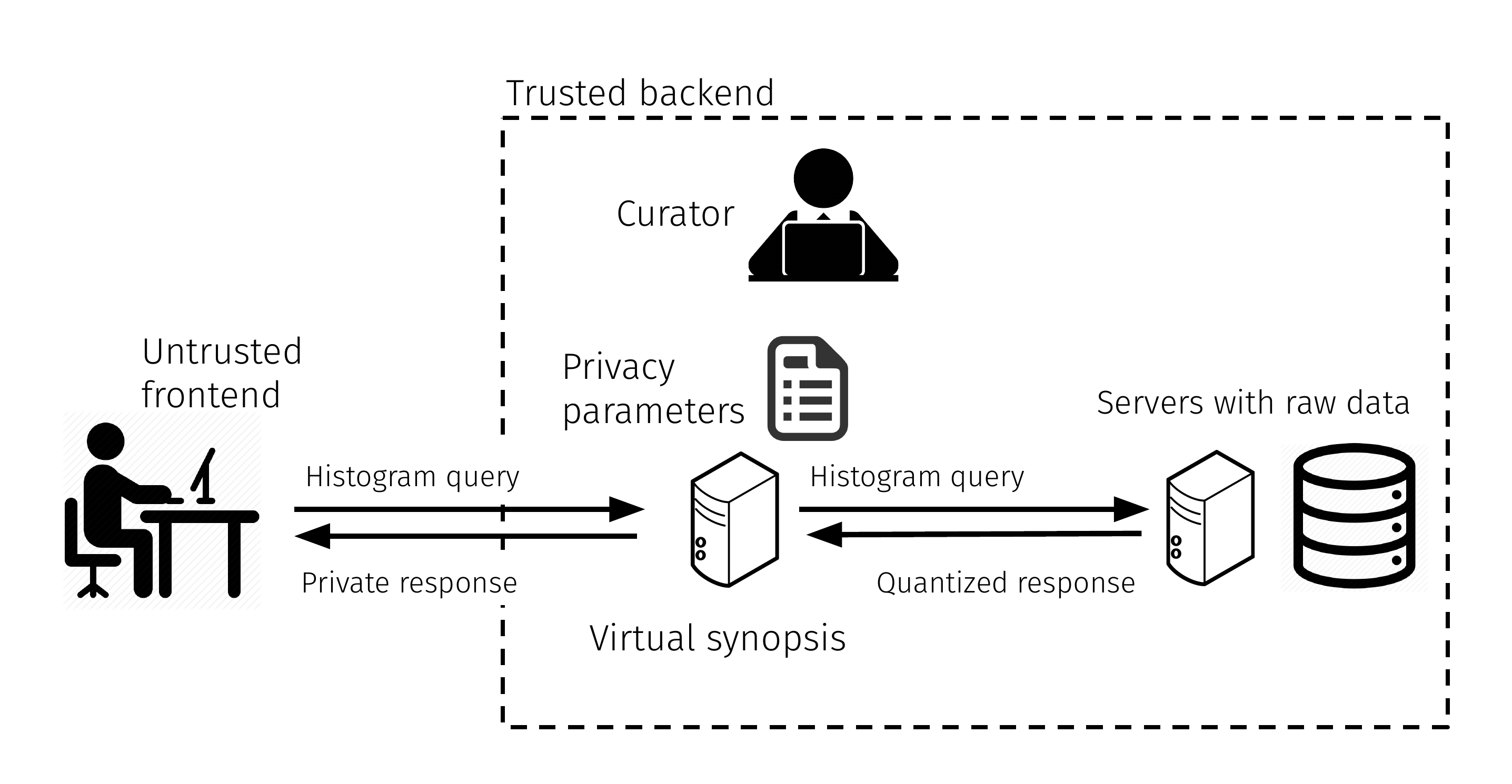}
    \setlength{\belowcaptionskip}{-20pt}
    \caption{Overlook architecture.\label{fig:overlook}}
  \end{center}
\end{figure}

Figure~\ref{fig:overlook} shows the architecture of \overlook.
A data analyst interacts with \overlook through a browser interface
that allows them to issue queries to the \overlook root node.
The root dispatches the query to the backend,
applies a privacy mechanism to the returned result,
and returns the private result to the user.

The raw data resides on a possibly distributed set of trusted servers;
a centralized root receives histogram queries and dispatches them to the servers.
The root also stores relevant privacy parameters used to compute
the private response to a histogram query.
The trusted data curator can make changes to these privacy parameters
until the dataset is published,
at which point the privacy parameters as well as the dataset must become immutable.
The untrusted data analyst can only access published data through
the private results of histogram queries.
The privacy \emph{parameters} are assumed to be public 
and visible to both the data curator and the data analyst.
The privacy parameters are discussed further in Section~\ref{sec:policies}.

\overlook primarily supports \emph{histogram queries}.
A histogram query over a column takes as input a set of disjoint buckets
and returns the number of data items that fall in each bucket.
In \overlook, these counts are perturbed with noise consistent with
a differentially-private mechanism,
described further in Section~\ref{sec:algo}.
\overlook allows users to issue an \emph{arbitrary number} of histogram queries
on privately-published data, with no privacy budget restriction.
Histogram queries are sufficient to support a large number of visualizations,
including histograms, heat maps, pie charts, trellis plots, and CDFs.
The visualizations supported in \overlook are described
in Section~\ref{sec:ui}.

In addtion to histogram queries,
\overlook's interface has support for other count-based queries,
such as the number of \code{NULL} values in a column.

\subsection{Threat model}

We assume untrusted users,
who can make an unbounded number of queries to the \overlook backend
through the \overlook UI.
Users may communicate with each other and
make queries in parallel or from multiple sessions.
Users cannot modify privacy parameters,
view raw data, or alter any secret key stored on the root.
Side-channel and denial-of-service attacks are out of scope in our work.

The data \emph{curator} is trusted
and has access to the raw data and privacy parameters.
The curator may not modify data or parameters once a dataset
has been published, as doing so would violate differential privacy.

The distributed backend is entirely trusted,
including the root server as well as 
servers hosting the raw data.

\subsection{User interface}

The data curator and data analyst both access
\overlook through interfaces that are extensions of the \hillview
data visualization system \cite{hillview},
which provides a browser interface for interacting with charts and data.
The curator UI is similar to the analyst's view,
except that the curator may additionally modify privacy parameters
and generate new histogram queries under the new parameters
prior to publishing the dataset.

The result of a histogram query is displayed as an interactive plot.
Additional queries can be made by zooming in using the mouse by selecting an interval,
which issues a new histogram query to the backend.

Note that, while the \emph{frontend} is an extension
to \hillview, \overlook can be used with any backend
that supports count queries. In Section~\ref{sec:implementation},
we describe one such alternate implementation using MySQL.

\subsubsection{Supported visualizations}
\label{sec:ui}

Overlook supports two main categories of private data summaries:
(1) histograms, and (2) counts of specific values, such as \code{NULL} values.
Multi-dimensional histograms encompass a number of useful visualizations,
including the traditional 1-dimensional histogram but also
heat maps, pie charts, trellis plots, and CDFs.

In addition, the user interface displays schema metadata
including standard values such as the column type,
but also the privacy policy associated with a column or group of columns.

\begin{figure}[t]
  \begin{center}
    \includegraphics[width=.8\columnwidth]{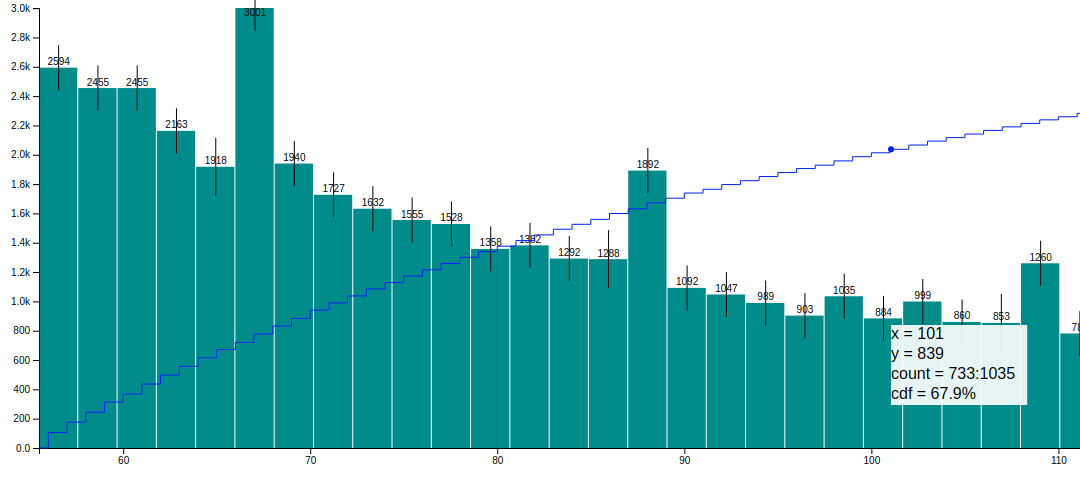}
    \setlength{\belowcaptionskip}{-20pt}
    \caption{Histogram plot with CDF curve overlaid. The data
      at the mouse position is shown in a semi-transparent white
      rectangle; notice that the bar size (count) is given as an
      interval, and confidence intervals are plotted for each bar.\label{fig:histogram}}
  \end{center}
\end{figure}

\begin{figure}[t]
  \begin{center}
    \includegraphics[width=.8\columnwidth]{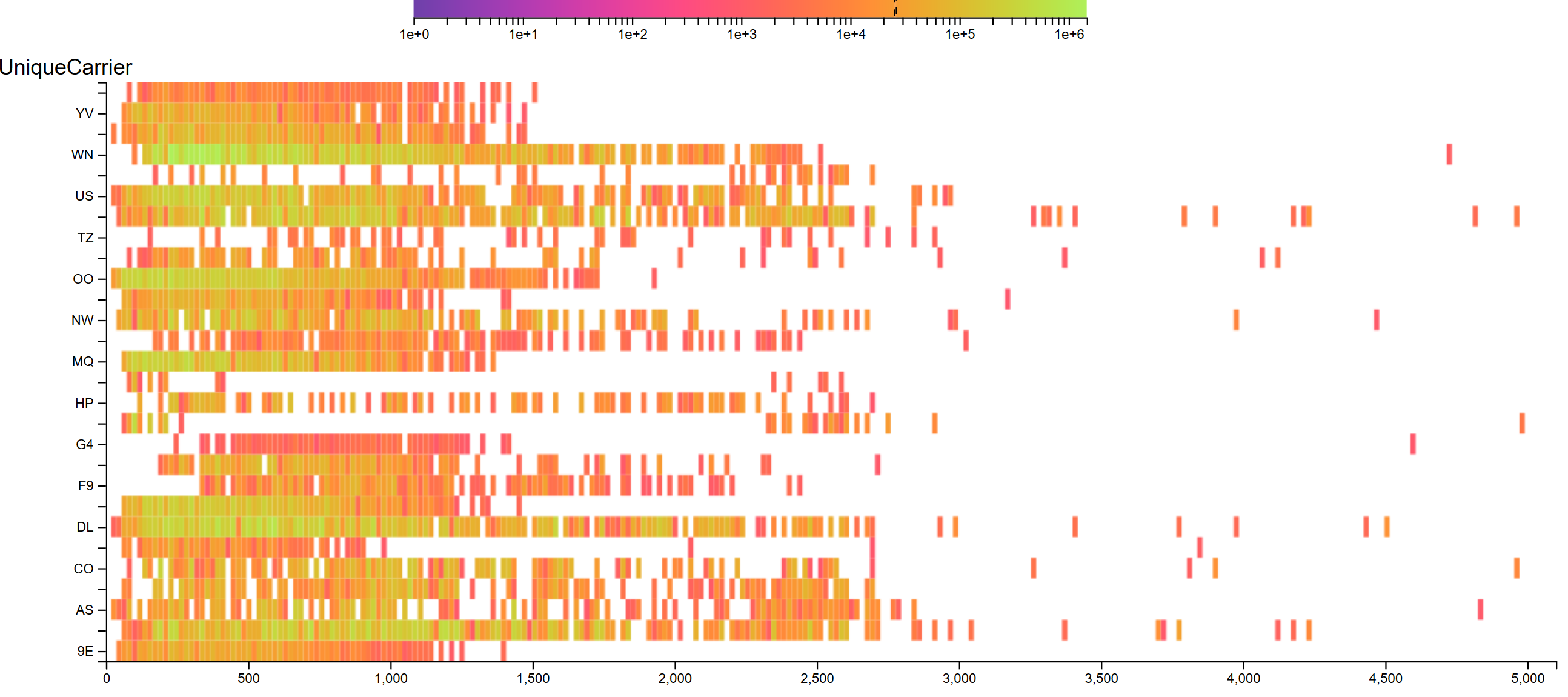}
    \setlength{\belowcaptionskip}{-20pt}
    \caption{Heatmap on two columns.  The color shows the count for
      each combination of values.  Values with low confidence are
      hidden, and the user can highlight with the mouse values within
      a specific range.\label{fig:heatmap}}
  \end{center}
\end{figure}

\paragraph{Histograms} \overlook's main primitive is a histogram query.
This primitive can be applied to create a variety of useful visualizations:
\begin{tightitemize}
  \item Histogram queries over a column (with numeric or categorical
    data).  The visual presentation can be a bar chart with confidence
    intervals, as shown in Figure~\ref{fig:histogram},
    or, for example, a pie chart,
    which emphasizes percentage of the whole that falls within each
    bucket. 
  \item Cumulative distributions functions (CDF) over a column
    (numeric or categorical).  CDF plots are always shown together with a histogram plot.
    Figure~\ref{fig:histogram} shows a
    histogram plot with an overlaid CDF curve.
  \item Histogram queries over a pair of columns, each of which can be
    either numeric or categorical.  This can be visually presented
    as a heat map as in Figure~\ref{fig:heatmap},
    or for example a trellis plot of 1-dimensional histograms.
\end{tightitemize}

One important feature of \overlook is that it displays estimates of \emph{uncertainty}
about the data. For 1-dimensional histograms, this is in the form of 99\%
confidence intervals.
The presentation of uncertainty is discussed further in Section~\ref{sec:experience}.

\paragraph{Counts} In addition to histograms, \overlook supports releasing
certain useful counts (``degenerate histograms'') privately:
\begin{tightitemize}
  \item In many views, the system displays information about the number
    of elements and the number of NULL values in a column.
  \item Distinct count queries: these estimate the number of distinct
    values in a column.
\end{tightitemize}

\subsection{Curator interface}
\label{sec:policies}

The data curator's job is to decide which
columns and pairs of columns will be released privately,
and to then decide the privacy level for each of those data releases.
\overlook's curator UI helps the data curator make these decisions.

For each set of columns that is to be released privately,
the curator must specify a corresponding \emph{privacy policy}.
This policy provides \overlook with information about public values
that can be used in the histogram as well as
parameters that are used to instantiate the privacy mechanism.
The curator's view of the \overlook UI allows
the curator to edit these policy settings and generate sample charts on a
dataset before it is published.
In this section, we describe the parameter settings the curator can choose in a privacy policy.

While specifying a policy for \emph{every} such histogram may be impractical,
\overlook provides some useful default values for the convenience of the curator.
While curators should be careful to choose parameters that are public and
independent of the data, we note that the curator's decisions may nevertheless
leak information because they are made based on the true underlying dataset.
Devising methods to add differential privacy to this kind of human-in-the-loop
parameter selection is an interesting avenue for future work.

\subsubsection{Privacy levels}

For each set of columns to release,
the curator specifies a corresponding value $\eps$ that denotes
the privacy level that should be used to release the column.
A smaller value of $\eps$ typically results in more privacy
at the cost of more noise in the private output.
The curator can explore many values of $\eps$ for each
set of columns before deciding which privacy level gives the
best tradeoff between data privacy and utility for potential analysts.

\subsubsection{Data ranges}

On the first histogram query a user makes for a column,
\overlook must return a histogram over a sensible range,
after which the user can zoom in to regions of interest.
A non-private visualization could compute the true minimum and
maximum values in the dataset and return a histogram over the full range.
Computing these values in a differentially-private manner
is a considerably more complex task \cite{DworkJ09}.
Instead, \overlook requries the data curator to specify publicly-visible
values for the initial minimum and maximum for each column
in the privacy policy.
As in prior work \cite{pinq, flex},
the curator must be careful to not choose values closely associated
with specific data points.

\subsubsection{Quantization}

In \overlook, the data curator must specify a public \emph{quantization},
or partitioning, of the data.
The need for this is twofold:
\begin{tightenumerate}
\item The synopsis used in \overlook, detailed in Section~\ref{sec:background},
  operates over a finite, enumerable data domain.
\item When displaying histograms on categorical data,
  \overlook uses the curator-specified bin boundaries as public \emph{labels}
  for the bins.
\end{tightenumerate}

To illustrate the second point,
consider a column that contains the names of patients in a hospital.
A non-private histogram might reveal specific names through the choice of bin labels.
In \overlook, a curator may set the quantization boundaries to
be the letters `A' through `Z',
so that the finest unit of aggregation is the first letter of the name,
and no individual's name is leaked in the published histogram.

The idea of public partitioning has been explored in prior systems \cite{pinq, flex},
but the partitioning in those systems has been left up to the data \emph{analyst}
to specify. In contrast, in \overlook,
the data \emph{curator} specifies the bin boundaries,
and can therefore choose a set of boundaries that is appropriate for the dataset
and provides good visual utility to analysts.

Importantly, \overlook never fully materializes a quantized version of
the dataset.
Instead, the quantization policy is expressed \emph{implicitly} as a function
that maps data points their quantized versions.

\section{Definitions and Data Model}\label{sec:background}

We consider a tabular dataset $X = \{x_1, \ldots, x_n\}$,
$x_i \in \A = \A_1 \times \ldots \times \A_d$,
which consists of $n$ rows and $d$ columns or attributes.
This dataset could be the result of a join on multiple tables
or a materialized view.\footnote{
  We note that a line of prior work \cite{ProserpioGM14, restricted-sensitivity, djoin, flex, privatesql}
  considers limiting the sensitivity of joins in differential privacy.
  \overlook assumes all released views are materialized and provides
  only per-row privacy in the materialized view,
  but, as \overlook operates over a standard database backend,
  these techniques could be incorporated in the future.
}
Each column $i$ takes values in the domain $\A_i$,
which must be finite and public
(which can be achieved implicitly by applying the privacy policy to each column
at query time).

Our goal is to answer queries from
some set $\Q$, where each $q \in \Q$ is a function $q: \A^n \rgta \Y$.
A mechanism for answering queries from $\Q$ is a randomized algorithm
$\M: \A^n \times \Q \rgta \Y$.
Following \cite{dp}, a mechanism $\M$ is $\eps$-differentially private if for any two
databases $X, X'$ which differ in a single row,
\[ \forall Y \subseteq \Y, \Pr[\M(X,q) \in Y] \leq
e^\eps\Pr[\M(X',q) \in Y]. \]

Intuitively, this means that adding, removing, or changing any one row of the dataset
will not change the probability of an event under the differentially-private mechanism
by more than a pre-specified multiplicative factor, $e^\eps$.
An extensive line of work has explored the construction of mechanisms
that achieve this guarantee \cite{dp-survey, salil-survey}.
The most common mechanisms add random noise to the raw result
according to the Laplace distribution.
In a histogram visualization, this manifests as random perturbations
to the counts for each bar.
Our challenge is then to choose a mechanism that gives
good visual utility to the user in spite of these random perturbations
(\S\ref{sec:hh}) and implement it efficiently (\S\ref{sec:algo}).

Importantly, throughout this section we assume that the mechanism operates over data
which has \emph{already} been quantized according to the privacy policy,
and therefore belongs to a finite and \emph{public} domain $\A$.
In practice, this quantization happens on demand, at query time;
the quantized dataset is not materialized.

\subsection{Query model}
\overlook primarily supports one- and two-dimensional histogram queries.\footnote{Larger dimensions can be accommodated easily, but histograms and heat maps are most relevant to the visualization setting.}

The basic building block of a histogram query is computing the count of the elements
that belong to a bucket; a bucket is in general a $k$-dimensional
rectangle.

For example, a 1-dimensional histogram query with $\ell$ buckets
specifies a column $i$ and a set of
$\ell+1$ bucket boundaries $h_0 < h_1 < \ldots < h_{\ell}$. 
The query returns a vector of $\ell$ counts, one for each interval $[h_i, h_{i+1})$.

  Note that, although the domain $\A_i$ must be finite and public,
  the bucket boundaries $h_i$ can be any value specified by the user.
  For example, if $A_i = \{0, 1, 2\}$,
  the user may query the range $[0.5, 1.5)$.
    In this case, \overlook would privately return the (noisy) count corresponding to the value 1
    (the only value in the quantized domain that falls in the query range).
    We are careful to note that this count corresponds to the data \emph{after} quantization.
    
    \overlook also supports releasing certain counts,
    such as counts of \code{NULL} or missing values,
    privately apart from the histograms.
    These can be made private simply
    by perturbing the count with noise distributed as $\Lap(1/\eps)$ \cite{dp}.
    The data curator must take into account the additional
    privacy cost of releasing these values.
    
\subsection{Synopsis mechanism}
\label{sec:hh}

Certain families of queries permit a special
type of mechanism that produces a private summary called a
synopsis. Such a mechanism can be decomposed into two stages:
\begin{tightenumerate}
  \item It releases a synopsis $S = \M_\eps(X)$ of the dataset, which
    is guaranteed to be $\eps$-differentially private, and is
    independent of the queries $q \in \Q$.
  \item On input $q \in \Q$, it computes an answer $S(q)$ using
    only the synopsis. Since the answer is computed by post-processing the synopsis, and
    post-processing does not leak privacy \cite{dp}, the answer is
    differentially private. Further, there is no limit to the number
    of queries that a user can submit.
\end{tightenumerate}

Given a column of a dataset over a finite, enumerable data domain $\A_i$ of size $m$,
one can build a histogram of $m$ buckets containing the
count of each element in the domain.
Making this histogram private na\"ively requires adding Laplace
noise with scale $\Lap(1/\eps)$ to each of $m$ buckets.
Answering an interval query of size $t$ then adds $t$
independent random Laplace variables to the result.
The error of such a mechanism scales as $\sqrt{t}$ \cite{binary-mechanism}.

\begin{figure}[t]
  \begin{center}
    \includegraphics[width=.8\columnwidth]{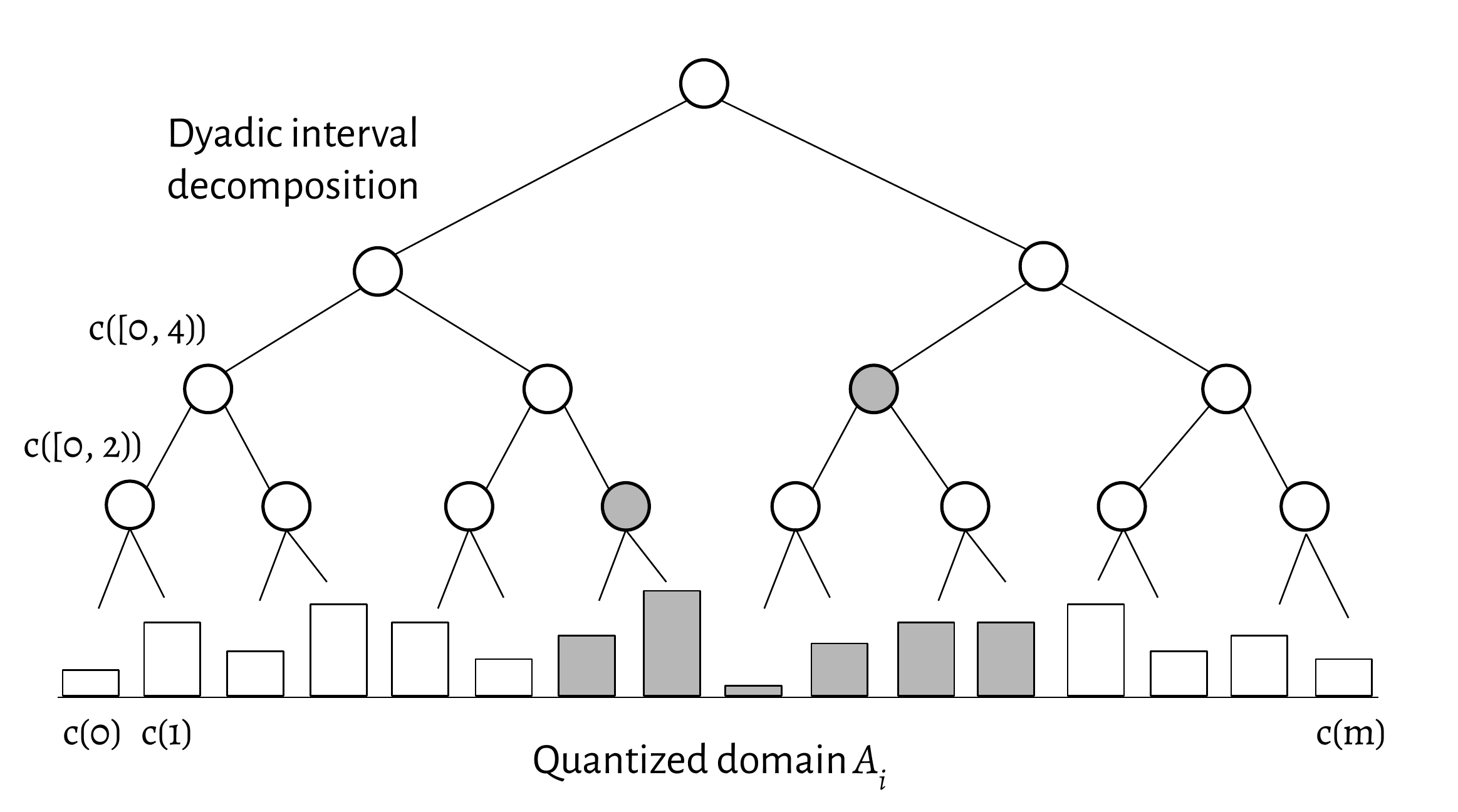}
    \setlength{\belowcaptionskip}{-20pt}
    \caption{Tree used in a hierarchical histogram with $b=2$.
      The tree for column $i$ is constructed for the quantized
      domain $A_i$. Each internal node in the tree corresponds to
      the count for a contiguous interval in the domain,
      and receives independent random Laplace noise.
      In this example, the grey internal nodes in the tree can be
      used to compute the count for the grey range in the domain. \label{fig:dyadic}}
  \end{center}
\end{figure}

However, adding this much noise to each query is suboptimal.
Instead, \overlook uses a mechanism called the \emph{\hh} \cite{Hay10, binary-mechanism},
also referred to as $\H_b$ in the literature.
The error of this mechanism scales as $\log_b^2(t)$ rather than $\sqrt{t}$.
At a high level, the \hh builds a \emph{tree}
such that nodes higher in the tree correspond to progressively larger contiguous intervals in the domain.
Each internal node of the tree corresponds to an interval of the histogram that is the
union of its $b$ children,
and the mechanism adds noise with scale $\Lap(\log_b(m)/\eps)$ to each internal node.
For such a tree with branching factor $b$,
an arbitrary interval of size $t$ can be computed by taking the union of only $(b-1)\log_b(t)$ nodes:
the number of noise variables now scales logarithmically,
rather than linearly, in the interval size.
Figure~\ref{fig:dyadic} shows such a tree with branching factor 2 (also called a ``dyadic'' tree).

\begin{figure}[t]
  \centering
    \includegraphics[width=.7\columnwidth]{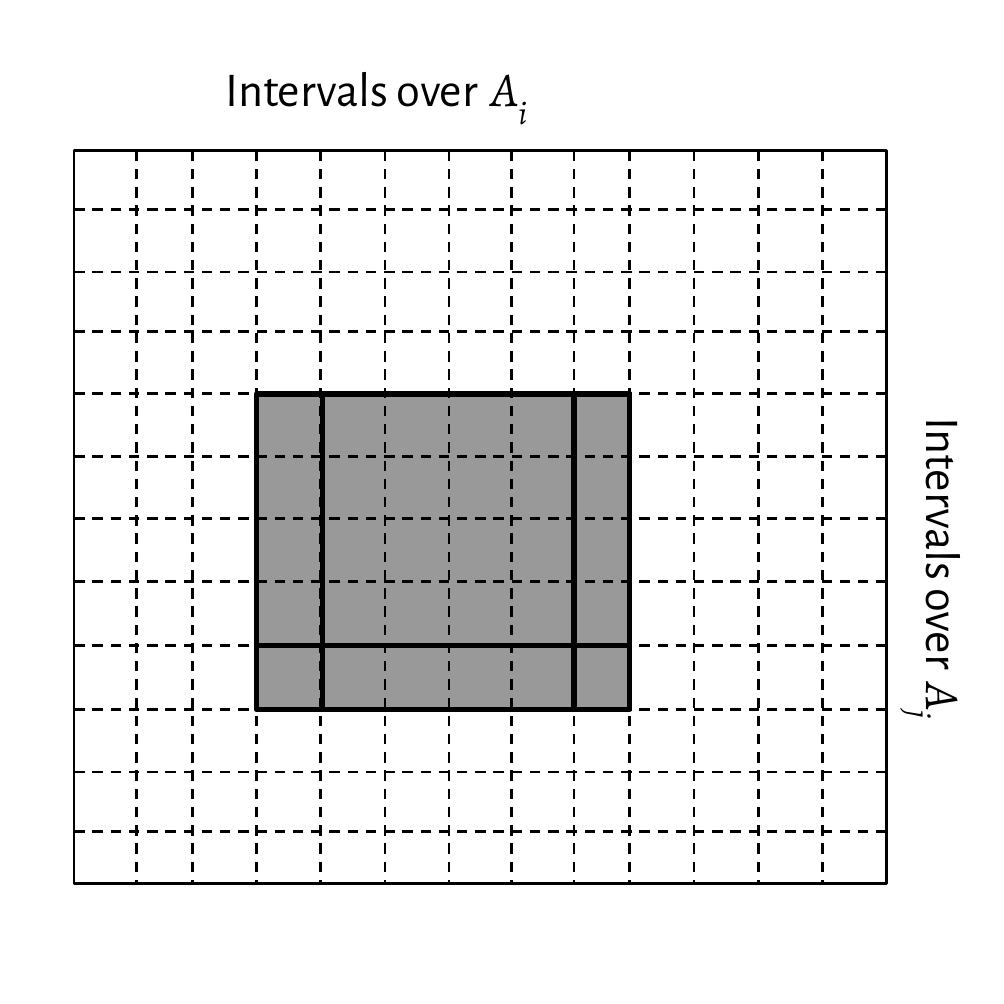}
    \setlength{\belowcaptionskip}{-20pt}
    \caption{Interval decomposition for a 2D histogram (heatmap).
      The decomposition in two dimensions is simply the Cartesian
      product of the decompositions in each dimension. \label{fig:2dnoise}}
\end{figure}

A multidimensional rectangle query can be computed
by taking the Cartesian product of its decomposition in each axis,
as illustrated in
Figure~\ref{fig:2dnoise}, for a heatmap over columns $i, j$.

\subsection{Discussion}
\paragraph{Why use hierarchical histograms}
Hierarchical histograms are just one of many mechanisms that have been proposed in
the literature (see the surveys of \cite{Hay10, QardajiYL13}). We choose
to implement hierarchical histograms for two reasons:

\begin{tightenumerate}
	
	\item {\em Data-obliviousness.} Hierarchical histogram mechanisms are
	oblivious to the data. They only need to know the quantization,
	which is public knowledge in our setting. This makes them
	particularly suitable for exploratory data analysis. We do not need
	any expensive processing of the data to compute quantization boundaries,
	which several data dependent mechanisms require.
	
	\item {\em Error guarantees.} A systematic comparative study of
	various data-dependent and data-indpendent mechanisms was performed
	by \cite{dpbench}. They found that for large datasets, the $\H_b$
	mechanism typically results in less error than any other
	mechanism. 
\end{tightenumerate}

\paragraph{Optimizing the shape of the tree}
Changing the topology of the tree yields different tradeoffs between
accuracy and privacy \cite{QardajiYL13}.  The
shallower the tree, the less the sensitivity of the resulting
synopsis, hence the less noise we need to add per node of the
tree. But then the number of nodes we need to sum might be large for
some ranges, which means those queries produce noisier results. The
tradeoffs between these parameters have been studied extensively in
\cite{QardajiYL13}. The best choice of $b$ depends on what type
of queries one wishes to optimize for. 

\paragraph{Setting privacy parameters}

The simplest option for the curator is to rely on the basic
composition theorem \cite[Lemma 2.3]{salil-survey} which states that
the privacy leakage adds up across mechanisms.  Hence, for range
queries of dimensionality $i$ the curator might specify a value
$\eps_i$, such that $\sum_{i \leq k} \eps_i = \eps$.  The curator may
then partition each $\eps_i$ (uniformly or otherwise) across the $d^i$
mechanisms that answer $i$-dimensional range queries.  \overlook's
privacy policy allows the curator to specify the value of $\eps_S$
for each set of columns $S$ independently.
A curator with greater expertise in differential privacy
may take advantage of advanced composition theorems~\cite{salil-survey}
to optimize the choice of $\eps$ for a table.

\section{Virtual Synopses}\label{sec:algo}

\begin{algorithm}[t]
  \hspace*{\algorithmicindent} \textbf{Input:} range $R$ for which to compute private count,
  branching factor $b$, domain size $m$, privacy level $\eps$,
  column index $i$, true count $c(R)$, PRF key $k$\\
  \hspace*{\algorithmicindent} \textbf{Output:} noisy count $\hat{c}(R)$
\begin{algorithmic}
  \caption{\overlook virtual synopsis.}
  \label{alg:synopsis}
\State Set scale $s = \lceil\log_b(m)\rceil / \eps$
\State Set $N(R) \leftarrow $ \textsc{B-adicDecomposition}$(R, b)$
\State Set $\hat{c}(R) = c(R)$
\For{$v \in N(R)$}
\State $\hat{c}(R) += \ell(F(k, (i, v)), s)$
\EndFor
\State \Return $\hat{c}(R)$
\end{algorithmic}
\end{algorithm}

\begin{algorithm}[t]
\textbf{Input:} interval $R = (r_l, r_r)$, $r_l \geq 0, r_r > r_l$; branching factor $b$\\
\textbf{Output:} nodes $N(R) = (v_1, \ldots, v_n)$ in tree corresponding to $R$,
indexed by (start, interval size)
\begin{algorithmic}
  \caption{Computing the $b$-adic decomposition for a range.}
  \label{alg:badic}
  \Function{B-adicDecomposition}{R, b}
  \State Set $N(R) \leftarrow \{\}$
  \State Set $L = r_l$
  \If{$|R| == 0$}
  \State\Return $N(R)$
  \EndIf
  \While{$L < r_r$}
  \State $p_L = -1$;
  \If{$r_l > 0$}
  \State $p_L = \lfloor \log_b(L)  \rfloor$ \Comment{Largest power of $b$ that divides $L$}
  \EndIf
  \State $p_s = \lfloor \log_b(r_r - L) \rfloor$ \Comment{Largest power of $b$ that fits in remaining interval}
  \State $p = p_L < 0 \text{ ? } p_S : \min(p_L, p_s)$
  \State nodeSize = $b^p$
  \State $N(R) = N(R) \cup \{(L, \mathrm{nodeSize})\}$
  \State $L += \mathrm{nodeSize}$
  \EndWhile
  \State\Return $N(R)$
  \EndFunction
\end{algorithmic}
\end{algorithm}

An important requirement for releasing a private synopsis
is that random noise is added once,
when the synopsis is constructed,
and must not be resampled on future queries to the synopsis.
For the \hh mechanism, this requirement na\"ively would mean
that \overlook would have to store a random sample for every
node in the synopsis tree,
a storage overhead that grows linearly in the size of the domain.

Our solution is to use a cryptographically secure
pseudo-random function (PRF) $F: (\K, \X) \to \Z$.
Informally, a PRF guarantees that, given a small random key $k$ in the key space $\K$,
$F(k, \cdot)$ will be indistinguishable from a truly random function
$R : \X \to \Z$ to a computationally-bounded adversary.
(See e.g. \cite{bonehshoup} for a formal definition.)

In our setting,
the inputs $\X$ to $F$ are nodes in the \hh synopsis for a given column,
each of which corresponds to a contiguous interval in the underlying domain.
$F$
takes as input a node index $v$, a column index $i$,\footnote{Multi-dimensional histograms are also assigned unique indexes, which can then be used with the PRF in the same manner;
  the general PRF for a node in a $d$-dimensional histogram with index $i$
  is $F(k, (i, v_1, \ldots, v_d))$.}
and the key $k$ associated with a table, and returns a
uniformly distributed random sample $F(k, (i, v))$,
which we then transform to a sample
from a Laplace variable $\ell(F(k, (i, v)), s)$ with the appropriate scale $s$.

Using the PRF, we are able to reduce the storage cost of the synopsis
from linear in the domain size to a small constant -- in fact, only the 32 bytes
required to store the key associated with a given table.

We incorporate the PRF into the \hh mechanism as follows.
A range $R$ can be decomposed into a minimal set $N(R)$ of internal nodes in the synopsis tree,
each of which corresponds to a sub-interval of $R$.
For each $v \in N(R)$, we can use the PRF to compute $\ell(F(k, (i, v)), s)$,
the random noise corresponding to that interval.
Then, if the true count in the interval is $c(R)$,
we can compute the private count for the interval
as $\hat{c}(R) = c(R) + \sum_{v \in N(R)}\ell(F(k, (i, v)), s)$.
(For a domain of size $m$, the scale $s$ is $\lceil \log_b(m) \rceil / \eps$.)

Algorithm~\ref{alg:synopsis} describes the synopsis algorithm in detail.
For completeness, we also describe the
algorithm for computing a $b$-adic decomposition of an interval
in Algorithm~\ref{alg:badic}.

Because the PRF is ultimately deterministic (though indistinguishable from random), 
this algorithm satisfies the requirement that queries
to the synopsis will not require resampling noise,
although we have not explicitly stored any samples.

This use of random number generators to save space is similar in spirit to
the use of such generators in streaming algorithms \cite{AMS96}.

\paragraph{Cryptographic security}
It is important to note that the PRF used to generate random samples must be \emph{cryptographically} secure.
If one can somehow reverse-engineer the generator and
compute $\ell(k, (i, v))$, then one can subtract it from the end result and
obtain the true count $c(v)$. A cryptographic PRF ensures
that, even if an adversary knows
$\ell(k, (i, v))$ for some $v$, perhaps because they know $c(v)$ as
auxillary information, the adversary still cannot efficiently
compute $\ell(k, (j, v'))$ for a new value and column.
This additionally requires that the key $k$ associated with a table must be stored securely
on the root node.

Section~\ref{sec:vs-impl} further describes the implementation of virtual synopses in \overlook.

\section{Implementation}\label{sec:implementation}

We have implemented Overlook by extending the open-source
Hillview~\cite{Budiu2019} visualization system.  However, to make the
case that a system like Overlook could be implemented as an agent
between any suitably powerful UI and a generic database, we have
extended Hillview with a custom back-end that generates SQL queries in
the SQL dialect of MySQL.  We then have added the Overlook
differential privacy layer on top of both these back-ends: the
Hillview in-memory database and MySQL.  For both cases the adaptations
required were minimal; the MySQL engine is completely unmodified.  We
describe them in the following sections.

In this section, we describe implementation details for the UI (\S~\ref{sec:ui-impl}),
privacy interposition layer (\S~\ref{sec:interposition}),
and \hillview and MySQL backends (\S~\ref{sec:backend-impl}).

\subsection{User interface}
\label{sec:ui-impl}
For data interaction and presentation we reuse the UI of \hillview \cite{hillview}.
This UI is written in TypeScript and runs in any modern web browser.
To support differentially private
visualizations we had to modify a few hundred lines of code.  The most
significant changes are (1) the data presentation of uncertainty
(confidence intervals) that is inherent in differentially private
results, and (2) the curator interface, which enables the curator to edit
the privacy parameters interactively.  We believe that displaying the
confidence intervals significantly enhances the usefulness of a
visualization tool.

\subsection{Privacy layer} 
\label{sec:interposition}

The privacy layer of Overlook is implemented in Java.  It is
sandwiched between the web server layer and the query generation and
execution layer.  When a new dataset is opened, the privacy layer
checks for the existence of related privacy metadata to decide whether
a data source should be treated as differentially private.

\subsubsection{PRFs for virtual synopses}
\label{sec:vs-impl}

The virtual synopsis described in Section~\ref{sec:algo}
uses a PRF to generate noise.
In practice, we use AES-256 as the PRF.
The root node stores one AES key \emph{per table},
which ensures that no two tables are released
using the same PRF.
In the same vein, columns and pairs of columns are labeled with
unique and immutable IDs so that no two synopses
within a table share random samples.

On receiving an interval query $[h_i, h_j)$,
  the root computes unique IDs of the nodes in the synopsis
  corresponding to this interval.
  To generate a new Laplace sample for an interval,
  the root uses the interval ID and column ID as input to AES 
  to generate random bits,
  which can then be transformed into a Laplace sample using
  standard methods for converting bits to doubles and then
  inverting the Laplace CDF.\footnote{
    We note that we do not currently implement the snapping mechanism
    described in \cite{ilya-dp},
    but this is not fundamental to our system design,
    and can be incorporated in the future.
  }

\subsubsection{Confidence intervals}

To approximate the $\alpha$-confidence interval for a sum of Laplace variables, 
we sample the corresponding distribution and return the $1-\alpha$ percentile value.
Na\"ively this operation would be performed for every bucket
on every histogram query.
However, we observe that our synopsis guarantees that every interval will be the sum of
at most $\log^d n$ random variables (for a histogram of dimension $d$,
where $n$ is the size of the domain).
Therefore, the confidence intervals once computed can be stored in a cache of size at most $\log^d n$.
Moreover, the confidence intervals are added as a postprocessing step
independent of the raw data,
and therefore need not be computed securely;
the cache can be shared across columns and tables.

\subsubsection{Privacy policies}

\overlook stores privacy policies in JSON format at the root node. 
Any user or curator can query the privacy policy,
but only the curator is allowed to modify it.

\subsubsection{Query rewriting}

The query rewriting layer receives queries from the UI and rewrites
them to operate on quantized data.  (The exception is the ``distinct
count'' query, which operates directly on the raw data, and not on the
quantized data.)  We give a concrete example about such query
rewriting in our implementation Section~\ref{sec:implementation},
where we describe the implementation of Overlook using a traditional
SQL database.  Recall that the quantization parameters for a column
are established by the data curator.  The quantization information
describes a range of intervals for the data values; data that falls
outside all the quantization intervals is treated as if it is a \code{NULL}
value.

\subsubsection{Adding noise to results} 

When the root receives the complete counts for the base histogram,
it queries the virtual synopsis for the noise to add to each bucket
and adds this noise to the histogram before returning it to the UI.

\subsection{Backends}
\label{sec:backend-impl}
Overlook's operation can be adapted to use any back-end that supports
a rich enough query language to compute standard histograms.  We
demonstrate this by describing how it operates over two different
back-ends: the Hillview back-end, and one that operates on top of
MySQL.  We are interested in highlighting the additional effort
required for adding privacy on top of an existing SQL-based query
engine.  In this section we describe how this is done for the case of
histogram queries.

\subsubsection{\hillview backend}

\hillview is a MapReduce-like distributed query engine 
that implements \emph{vizketches} --
mergeable sketches for visualization.
It implements a number of data-parallel aggregation tasks
suitable for visualization, including those used in \overlook.
\hillview is described in additional detail in \cite{Budiu2019}.

The only change required to support privacy-related processing is to adapt all the
existing sketches to first quantize the columns that they operate on
according to the appropriate privacy policy.
No other changes were required in the backend.
This change amounts to less than 10 lines
of code in each sketch.

\subsubsection{MySQL backend}

\overlook can interface with unmodified, existing database backends;
we have implemented one such backend in MySQL.
In this section, we describe some of the queries implemented in order
to support the \overlook UI.

\paragraph{Numeric histograms}

Consider a user request to display a (non-private) histogram of the
data in a column \code{C} as a histogram with \code{b} buckets.  Let
us assume first that \code{C} is a numeric column in table \code{t}.
This kind of visualization is executed using SQL in two stages: (1)
the range of the data in the column is computed, and (2) the histogram
is built.  To obtain the range of the data we generate the following
query:

\begin{lstlisting}[language=SQL]
SELECT min(C), max(C), count(*), count(C)
FROM t
\end{lstlisting}

This query computes the minimum and maximum values in column
\code{C}, and also the number of non-null elements and the number of
total elements.  

The UI receives these parameters and decides on a range
\code{l}--\code{r} of data and on a number of buckets \code{b} to
display (in some cases the UI does not need to issue any other query,
for example when all elements are \code{NULL}, or when \code{l=r}). 
The query to compute a histogram is written as:

\begin{lstlisting}[language=SQL]
SELECT bucket, COUNT(bucket) FROM (
  SELECT CAST(FLOOR((C - l) * scale)
     AS UNSIGNED) AS bucket
  FROM t
  WHERE C between l AND r)
GROUP BY bucket
\end{lstlisting}

\noindent \code{scale=b/(r-l)} is computed statically before the
query is generated.

\paragraph{Quantized data view}

Since all private queries operate over quantized data, one option is
to pre-compute and materialize a view where all columns are quantized
using the curator-specified quantization intervals.  Such a query can
be generated automatically by the system once the privacy policy has
been set.  For example, to create a view \code{QV} of a table with a
single numeric column \code{C} with equal-sized quantization intervals
of size \code{g} between \code{qmin} and \code{qmax} one can issue the
following query:

\begin{lstlisting}
CREATE view QV as
  (SELECT qmin + FLOOR((C-qmin)/g)*g AS C
   FROM t WHERE C between qmin AND qmax)
\end{lstlisting}

\paragraph{Private numeric histograms}

For the case of a private numeric column the general flow is very much
as described in the previous section; there are two changes: (1) the
query is executed over the quantized view, and (2) after the histogram
is computed noise is added to each bucket.  Let us assume that we are
quantizing the data to be within the range \code{qmin} and \code{qmax}
with a granularity~\code{g}.

The complete query that is executed is:

\begin{lstlisting}[language=SQL]
-- compute histogram  
SELECT bucket, COUNT(bucket) FROM (
  -- compute buckets  
  SELECT CAST(FLOOR((C - l) * scale)
     AS UNSIGNED) AS bucket
  FROM QV -- quantized view
  WHERE C between l AND r)
GROUP BY bucket
\end{lstlisting}

\paragraph{String histograms}

Computing (non-private) histograms over a categorical column is a bit
more involved because the UI never displays a large number of
histogram buckets (more than can be shown on the screen).  The first
step in computing a histogram over a string column involves computing
a set of \emph{distinct quantiles} over the column.  For example, if
the screen can accommodate 50 columns, then the UI will first issue a
query to sort the distinct values in the column and extract 50
equi-distant values from the sorted set.  These 50 values will be used
as histogram bucket boundaries.  If the column has fewer than 50
distinct values then all values will be used as distinct bucket
boundaries.

\begin{lstlisting}[language=SQL]
SELECT DISTINCT BINARY C AS C FROM t
ORDER BY BINARY C
\end{lstlisting}

(One has to be careful with the sorting and comparison order: these
have to be consistent between the code that computes the buckets and
the database code that performs comparisons and sorting.  In our case
we had to prevent MySQL from doing default case-insensitive string
comparisons in order to obtain consistent results --- this is why we
used the \code{BINARY} keyword.  We will omit it from the subsequent
queries.)

To compute a histogram quickly over a set of explicit string buckets
the Java code generates an explicit binary search tree using nested
SQL \code{IF} expressions.  For example, to build a histogram with
buckets separated by strings `A', `G', `M', and `Z' it generates the
following query:

\begin{lstlisting}[language=SQL]
SELECT bucket, count(bucket)
FROM (
  SELECT (IF(C<`G',0,IF(C<`M',1,2))) AS bucket
  FROM t
  WHERE C BETWEEN `A' AND `Z')
GROUP BY bucket  
\end{lstlisting}

\paragraph{Private string histograms}

Finally, computing private histograms requires modifying the query for
string histograms in a way similar to numeric private histograms, by
quantizing the data in the column first.  A quantization policy for a
string column is given essentially by a sorted list of strings.  The
quantization query also makes use of a binary search tree.  Let's
assume that our quantization boundaries are `A', `F', `N', `O' and
`Z'.  The quantization query is:

\begin{lstlisting}[language=SQL]
CREATE view QV as  
  (SELECT IF(C<`N', IF(C<`F', `A', `F'),
                 IF(C<`O', `N', `O')) AS C
   FROM t
   WHERE C BETWEEN `A' AND `Z')
\end{lstlisting}

The query to compute a histogram over a quantized view is the
composition of these two queries:

\begin{lstlisting}[language=SQL]
SELECT bucket, count(bucket)
FROM (
  SELECT (IF(C<`G',0,IF(C<`M',1,2))) AS bucket
  FROM QV -- query the quantized view
  WHERE C BETWEEN `A' AND `Z')
GROUP BY bucket  
\end{lstlisting}

\section{Experience}\label{sec:experience}

Since we have built \overlook on top of the UI of an existing
visualization tool, we can make a direct comparison of the user
experience for traditional and differentially-private visualization.
In this section, we describe some notable differences between
these user experiences.

\paragraph{Browsing individual data items} The most conspicuous difference is that many operations that are
natural in a normal visualization are unavailable when doing
differentially-private visualization.  For example, enumerating the
rows of a table is something that cannot be done in a
differentially-private way.  Traditional visualization systems can be
used for two purposes: detecting trends and identifying outliers.  A
differentially-private visualization system can only be used for the
first purpose: differential privacy masks rare events.

\paragraph{Displaying uncertain values} A second difference is that all counts that are displayed in a
differentially-private visualization are noisy.  This can be
interpreted as displaying a value with uncertain range.  Although
there is substantial work on the visualization of uncertain values,
from our experience the interpretation of confidence intervals
requires a sophisticated understanding from users.
While confidence intervals are a useful tool to visualize uncertainty,
they do not prevent spurious high counts
that users might confuse for signal.
This phenomenon has already been observed
by~\cite{zhang-dpw16}.

\paragraph{Uncertainty in heat maps}
\begin{figure}[t]
  \includegraphics[width=\columnwidth]{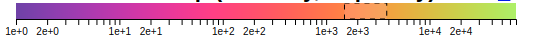}
    \setlength{\belowcaptionskip}{-10pt}
  \caption{An example of a confidence interval (dotted box)
    overlaid on a heat map legend. When a user hovers over a heat map cell,
    \overlook highlights the confidence interval for the box on the legend.}
  \label{fig:confidence-legend}
\end{figure}

\begin{figure*}[t]
  \centering
  \begin{subfigure}{.24\textwidth}
    \includegraphics[width=\textwidth]{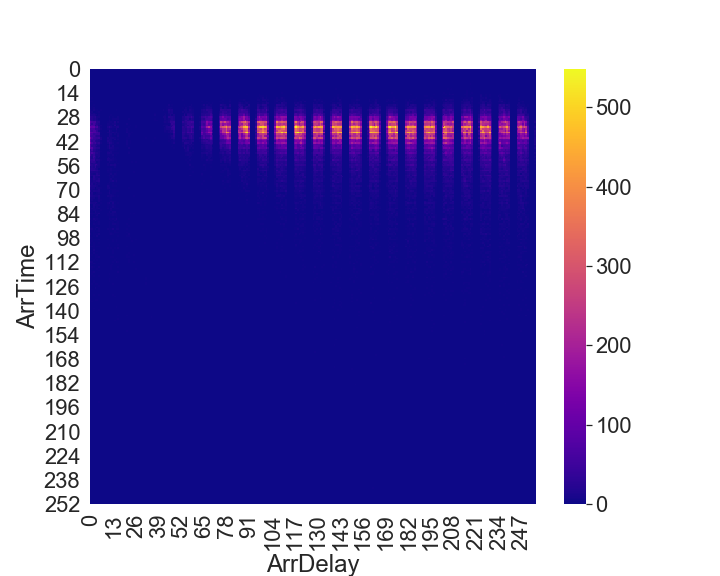}
    \caption{}
    \label{fig:rawhist}
  \end{subfigure}
  \begin{subfigure}{.24\textwidth}
    \includegraphics[width=\textwidth]{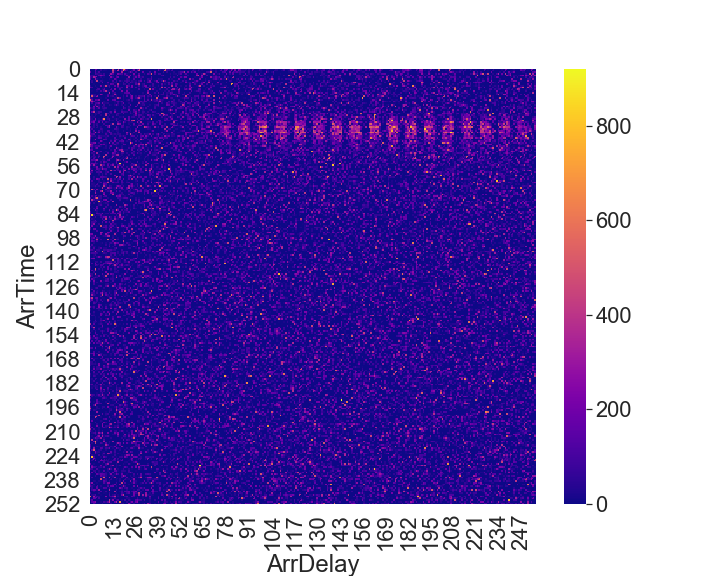}
    \caption{}
    \label{fig:noisyhist}
  \end{subfigure}
  \begin{subfigure}{.24\textwidth}
    \includegraphics[width=\textwidth]{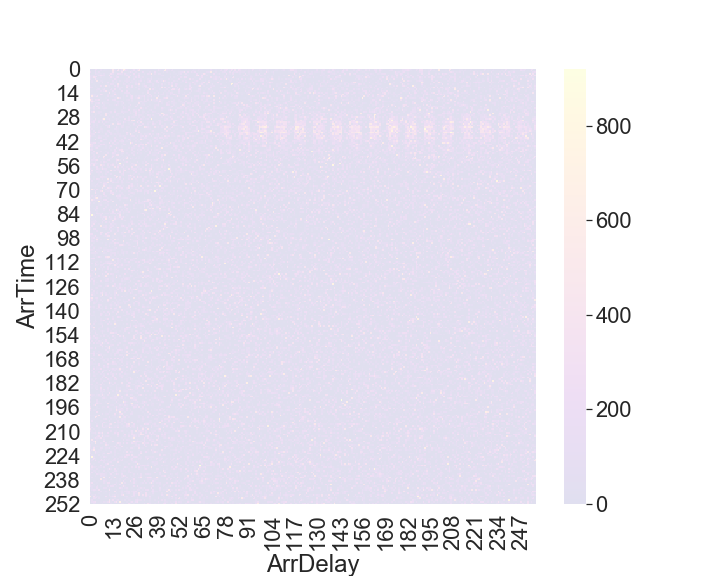}
    \caption{}
    \label{fig:256bins}
  \end{subfigure}
  \begin{subfigure}{.24\textwidth}
    \includegraphics[width=\textwidth]{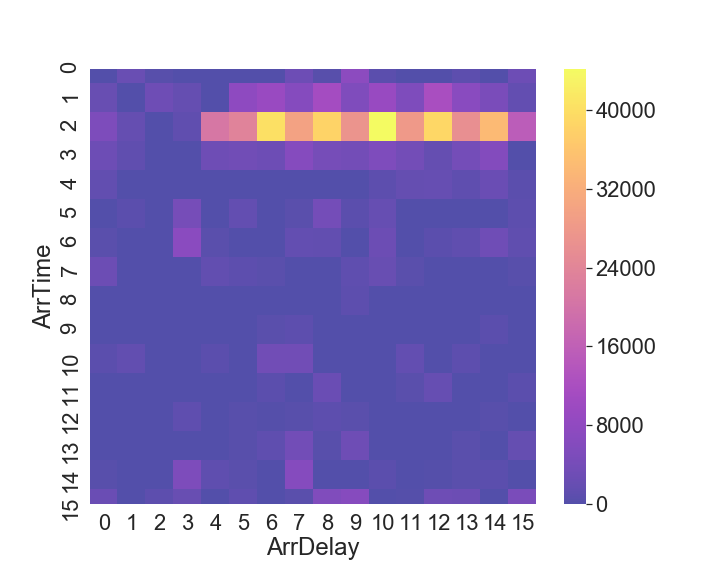}
    \caption{}
    \label{fig:16bins}
  \end{subfigure}
  \caption{Whitening to convey uncertainty.
    The color scale of the heat map conveys the \emph{value} in each cell;
    whitening adds an additional axis of color that can be used to convey uncertainty.
    (\ref{fig:rawhist}) Raw histogram with no noise or whitening.
    (\ref{fig:noisyhist}) Histogram with noise and no color adjustment.
    (\ref{fig:256bins}) Whitening added to raw histogram on fine-grained bins.
      Each bin has a small count relative to the standard deviation
      of the added noise, which is conveyed through the whiteness of the chart.
    (\ref{fig:16bins}) Whitening added to raw histogram on coarse-grained bins.
      Each bin now has a larger count relative to the standard deviation
      of the data, so less whitening is applied.
  }
  \label{fig:whiteness}
\end{figure*}

While uncertainty for histograms can intuitively be presented
as a range around each count,
it is less clear how uncertainty should be displayed in a heat map.
We prototyped multiple possible solutions to this problem.

Figure~\ref{fig:confidence-legend} shows a heat map legend
in \overlook that displays a confidence interval.
When the user hovers over a cell, \overlook highlights the
confidence interval in the legend.

In addition, however, we would like to visually
convey the confidence in each cell on the chart itself.
One way to achieve this, shown in Figure~\ref{fig:heatmap},
is to suppress any values whose count is smaller than
a multiplicative factor of its confidence interval.
As a result, only counts that are likely to be informative are displayed.

Another idea is to use the \emph{whiteness} of the
image to convey uncertainty.
In figure~\ref{fig:whiteness}, we demonstrate a prototype of such a plot.
The raw color scale is used to convey the count in each cell,
and the whiteness provides an additional dimension that can
be used to convey the amount of certainty a user should have in the visualization.

\begin{figure}[t]
  \centering
  \begin{subfigure}{.8\columnwidth}
    \includegraphics[width=\columnwidth, height=.8in]{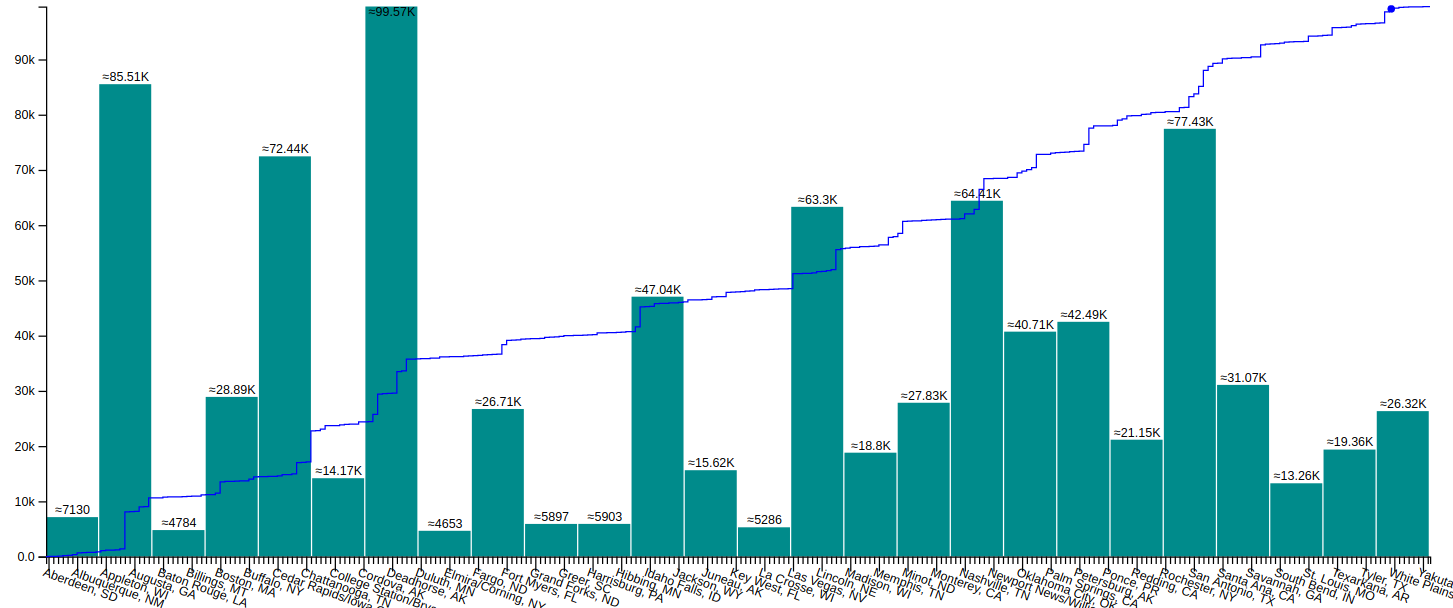}
    \label{fig:cities-full}
  \end{subfigure}
  \begin{subfigure}{.8\columnwidth}
    \includegraphics[width=\columnwidth, height=.8in]{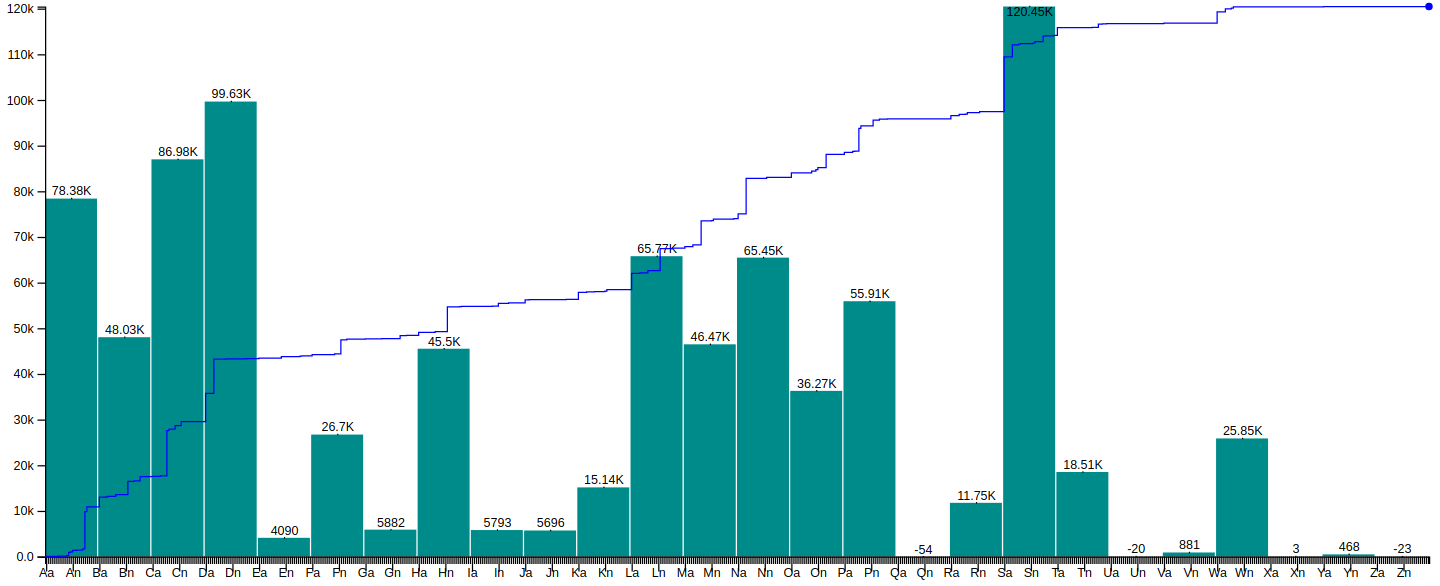}
    \label{fig:cities-2letter}
  \end{subfigure}
  \setlength{\belowcaptionskip}{-20pt}
  \caption{Histogram with up to 26 buckets of a set of cities
    quantized in different ways: (upper) public histogram
    with adaptively chosen bin boundaries; (lower)
    quantization fixed to two-letter boundaries.
    The shape of the histogram changes according to the choice of bins.
\label{fig:quantization}}
\end{figure}

\paragraph{Quantization intervals} The quantization intervals, especially for categorical data, have a
huge impact on the information that is conveyed to the user.  As an
example, we show in Figure~\ref{fig:quantization} several histograms
of the exact same dataset on the column ``cities'' with different
quantization intervals.  The first histogram has on the X axis only
the cities that actually appear in the dataset, sorted alphabetically,
so the distribution of cities into buckets is quite different.  The
last histogram allows the user to zoom-in further and explore the
distribution of data for each letter pair; the additional structure is
visible in the CDF which is more fine-grained (it has $26^2$ steps
instead of just $26$).

\paragraph{Query resolution} The amount of noise added to a histogram/heatmap bucket $R$ on columns
$S$ depends on two primary factors: the extent of the bucket (the set
of quantization intervals that fall in $R$), and the $\epsilon_S$
privacy budget allocated for the set $S$ of columns.  The noise does
not depend on the actual data distribution; however, the
\emph{relative} noise added does depend on the number of data items
that falls into the bucket $R$.  So there is a trade-off between the
resolution of the query and its precision: if we make buckets smaller,
we can potentially see more detail in the data, but the relative noise
will be higher.  If we make buckets larger we lose the resolution but
we gain precision.  There is no obvious choice in this trade-off,
since it depends very much on the data distribution.  This is a
trade-off that the data curator can explore and to some degree control
by choosing the $\epsilon_S$ and the quantization intervals for each
column.

\paragraph{Outliers or sentinel values} In one database we have encountered a date column which was using a
value year of 9999 to indicate that an event has not happened yet.
\overlook in general releases counts for \code{NULL} values separately 
from the \hh synopsis,
as \code{NULL} will not be part of any range query.
In contrast, this sentinel value would be na\"ively included in the
displayed histogram if the specified public date range included values up to 9999.

\section{Evaluation}\label{sec:evaluation}

In this section, we evaluate the design decisions made in \overlook
to support our claims that the system:
\begin{tightitemize}
\item allows data curators to quickly explore parameter settings for synopses
  before data release (\S~\ref{sec:synopsis-time}),
\item implements a synopsis that provides accuracy comparable
  to state-of-the-art methods (\S~\ref{sec:synopsis-acc}),
\item achieves significantly lower storage cost than the synopses
  implemented in prior systems
  through the use of a pseudorandom function (\S~\ref{sec:storage}), and
\item retains the scaling properties of the underlying distributed system
  with low performance overhead from privacy (\S~\ref{sec:scaling}).
\end{tightitemize}

In addition, we demonstrate that, in the visualization setting,
the error induced by differential privacy can be \emph{smaller than a pixel}
on the screen -- so that, with high probability,
the user loses no utility compared to the raw visualization (\S~\ref{sec:pixel-error}).

\paragraph{Evaluation setup} Local experiments were run on a machine
with 16 GB of memory and 4 cores using an Intel i7 processor.
Cloud experiments were run on an Amazon EC2 cluster of 15 machines
with 8 GB of RAM and 2 cores each.
The \hillview backend uses Java 8.

\subsection{Synopsis generation overhead}
\label{sec:synopsis-time}

One benefit of \overlook is that it allows the data curator to quickly
explore privacy parameters for the data before it is released.
In particular, the data curator might change the data quantization
or privacy budget $\eps$ for any column.
Generating example visualizations with the new parameters then requires
recomputing the underlying synopsis.

We use DPBench~\cite{dpbench} to evaluate
the time required to generate a synopsis with the \hh mechanism
against the time required for comparable synopses.
We stress that these times are not trivially comparable as
DPBench is primarily an accuracy benchmark that is not optimized for performance.

\begin{figure}[t]
  \centering
  \begin{subfigure}{.49\columnwidth}
    \includegraphics[width=\columnwidth]{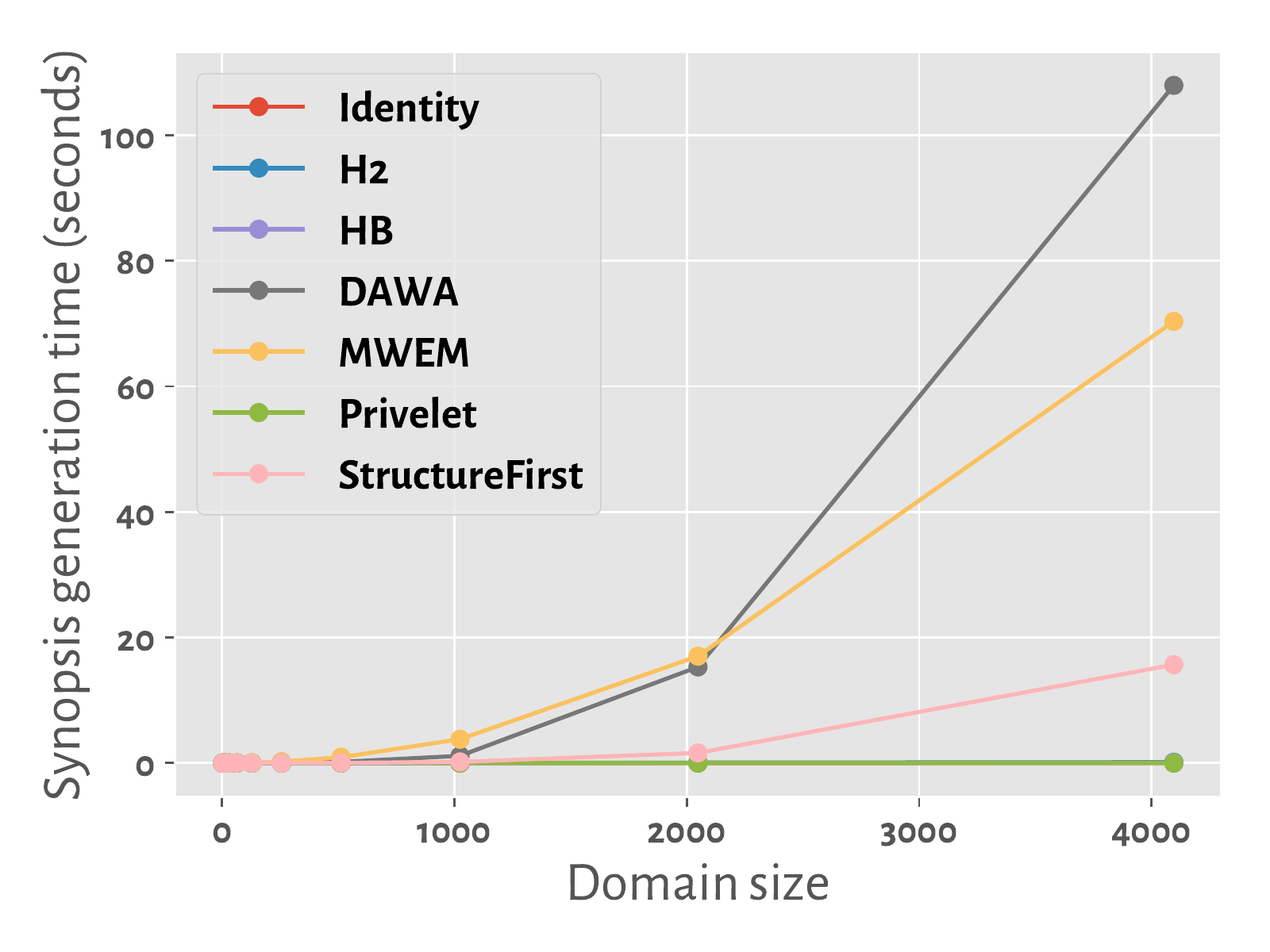}
    \caption{Time required to generate synopses as the domain size increases.
    \label{fig:synopsis-large}}
  \end{subfigure}\hfill%
  \begin{subfigure}{.49\columnwidth}
    \includegraphics[width=\columnwidth]{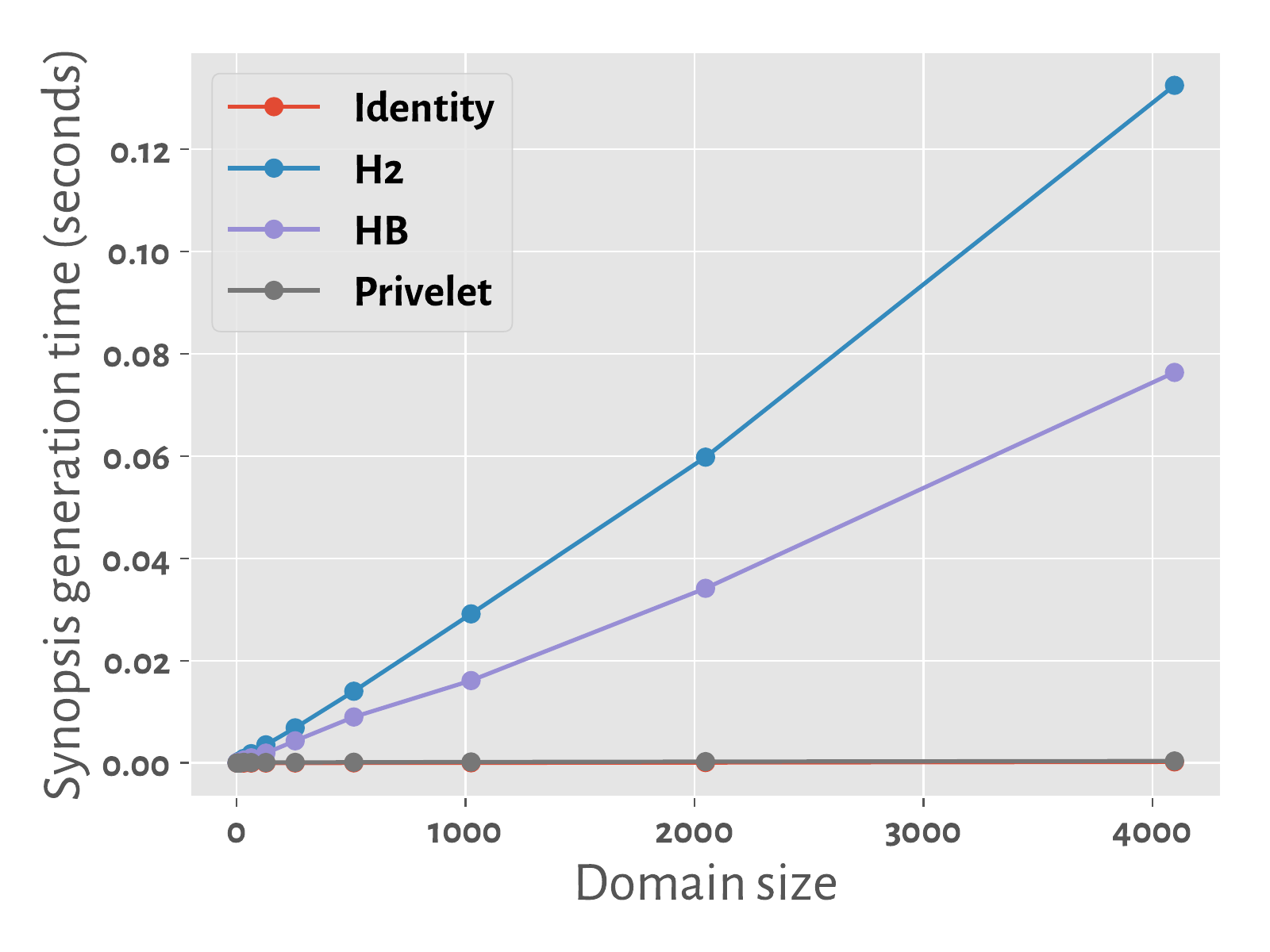}
    \caption{Generation time for faster synopses.
    \label{fig:synopsis-small}}
  \end{subfigure}
    \setlength{\belowcaptionskip}{-15pt}
  \caption{Time required to generate synopses using various mechanisms,
    benchmarked using DPBench. MWEM and DAWA dominate in the first plot;
    the second plot shows that generating hierarchical histograms scales
    in the domain size, when not using \overlook's PRF-based construction.
    \label{fig:synopsis-time}}
\end{figure}

We evaluate each method on a one-dimensional all-zeros dataset of increasing size,
on a workload of all intervals (the workload that \overlook targets).
These times do \emph{not} include the additional
time required to compute the base histogram of counts
over which the synopses are computed.
We evaluate seven mechanisms in the literature:
the baseline ``identity'' mechanism \cite{dp},
the binary \hh \cite{binary-mechanism, Hay10},
the \hh with adaptive branching \cite{Hay10},
DAWA \cite{dawa}, MWEM \cite{mwem},
Privelet \cite{privelet}, and StructureFirst \cite{xu2013differentially}.

Figure~\ref{fig:synopsis-time} shows the results of the benchmark.
Figure~\ref{fig:synopsis-large} shows that MWEM and DAWA
are by far the most expensive algorithms, followed by StructureFirst.
The remaining algorithms run in under one second, so we
plot these separately in Figure~\ref{fig:synopsis-small}.
While the time required for the hierarchical mechanisms scales
linearly in the data size, they are still considerably less expensive to compute
than more complicated workload-aware synopses.

In fact, \overlook itself does not instantiate the synopsis
and computes the noisy values on the fly,
so the synopsis adds no precomputation overhead
but does add some overhead at \emph{query} time.
However, this overhead scales with the number of \emph{histogram buckets} in the query, 
rather than the size of the domain or dataset.
We benchmark \overlook query overhead in Section~\ref{sec:scaling}.

\subsection{Synopsis accuracy}
\label{sec:synopsis-acc}

\begin{figure*}[t]
  \begin{subfigure}{\columnwidth}
    \includegraphics[width=\columnwidth, height=1.8in]{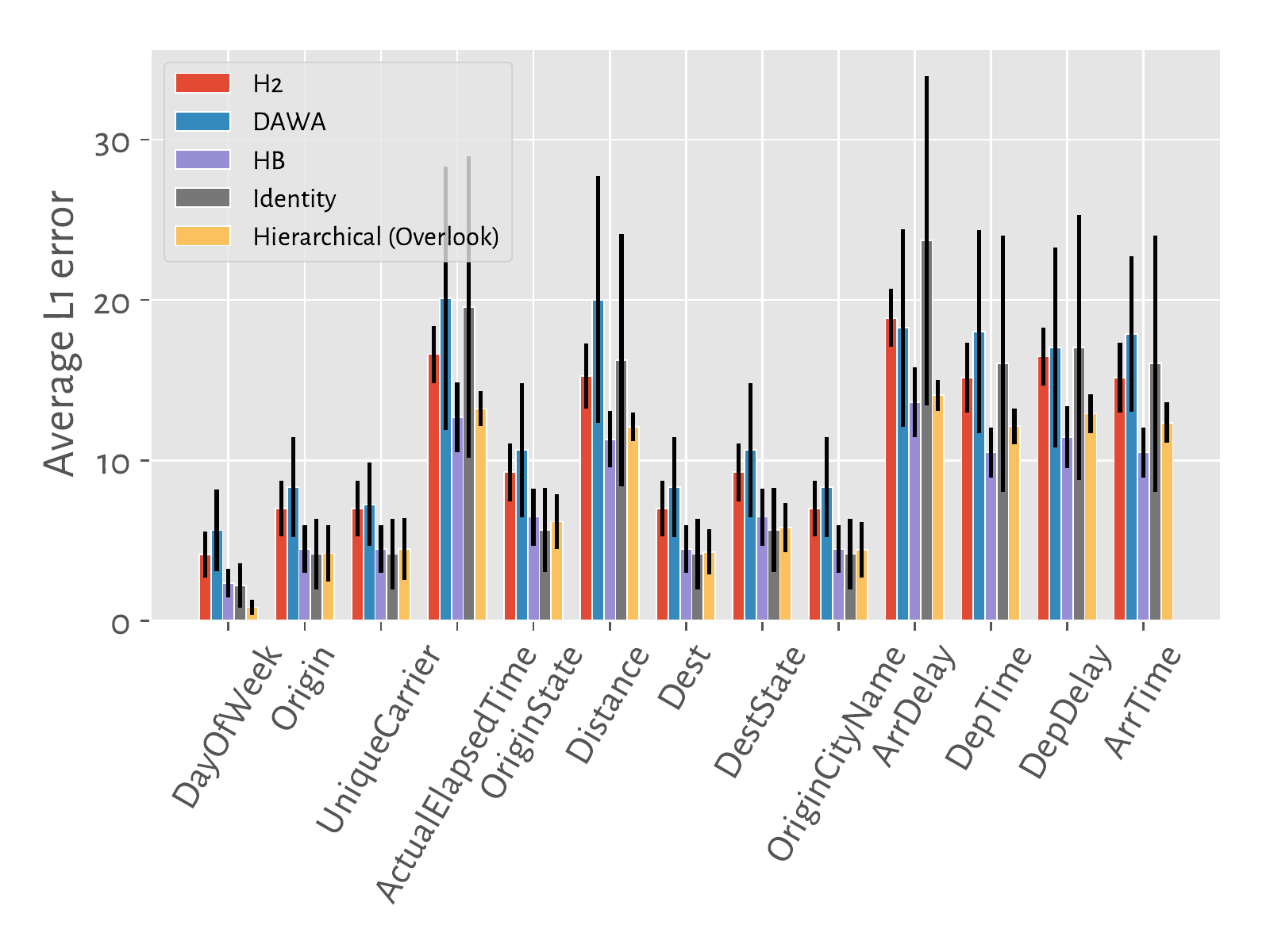}
    \caption{Histogram accuracy. Most mechanisms perform comparably on the flights dataset.
      \label{fig:accuracy-hist}}
  \end{subfigure}\hfill%
  \begin{subfigure}{\columnwidth}
    \includegraphics[width=\columnwidth, height=1.8in]{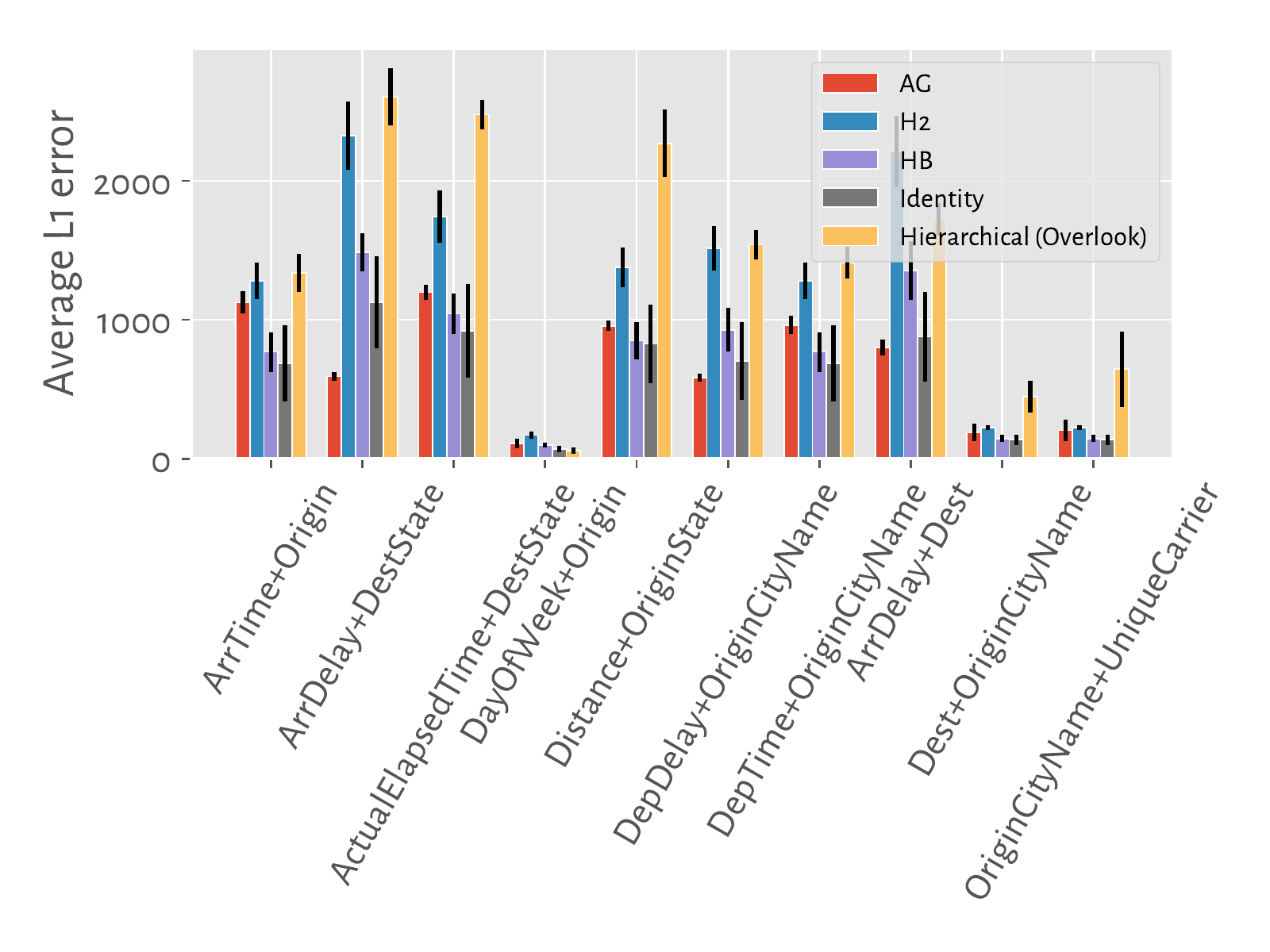}
    \caption{Heatmap accuracy. The baseline identity mechanism outperforms all others;
      the hierarchical mechanism nevertheless achieves reasonable accuracy.
      \label{fig:accuracy-heat}}
  \end{subfigure}
    \setlength{\belowcaptionskip}{-15pt}
  \caption{$\ell_1$ error of Overlook mechanism on the U.S. flights dataset on 5000 randomly-sampled
    queries per column or pair of columns.
    \label{fig:accuracy}}
\end{figure*}

\overlook uses hierarchical histograms as the underlying synopsis.
The primary benefit of the hierarchical histogram is that
the error scales logarithmically, rather than linearly, in the size of the underlying dataset.
However, more complex optimization procedures \cite{matrix-mechanism, dawa}
may yield even better accuracy.

In this section, we demonstrate empirically that \overlook's histogram
mechanism achieves utility comparable to that of state-of-the-art synopses.
These results are supported by prior work \cite{QardajiYL13, qardaji2013differentially, Hay10}
that also investigates the empirical accuracy of hierarchical histograms.
More complex methods work well for skewed or restricted query workloads,
but \overlook benefits from simple mechanisms because it aims to support a very general
set of range queries.

We benchmark accuracy on a dataset of 20 years of U.S. flights \cite{ontime}.
This dataset contains both numeric and categorical columns
over a range of data distributions and domain sizes (varying from 7 to over 4000).
Figure~\ref{fig:accuracy} shows results for histograms on all columns
and heat maps on a selection of columns.
For each bar, we sample 5000 random intervals or rectangles
and compute the $\ell_1$ distance between the vector of true counts for all samples
and the vector of noisy counts returned by the mechanism.

The key takeaway from these figures is that
the \hh mechanism has comparable accuracy to
mechanisms that perform more complex, workload-specific optimization
on the random-intervals workload.
As noted in \cite{QardajiYL13}, the benefits of this mechanism
decrease as the dimension increases.
Adaptively choosing which mechanism to use for a given visualization
may be a direction for future work.

\subsection{Synopsis memory overhead}
\label{sec:storage}

\overlook's synopsis mechanism has only 32 bytes of memory overhead
required to store the AES secret key.

The synopsis mechanisms implemented by DPBench are \emph{consistent} mechanisms.
These take as input a histogram over the elements of the domain
and output a synthetic histogram as the synopsis over which
all subsequent queries are run.
The size of the synopsis is therefore proportional
to the size of the data domain for a histogram,
and grows exponentially in the number of dimensions.
(In particular, if the data is small or especially sparse in the domain,
the size may be considerably larger than necessary to represent the data.)

We note, that these mechanisms may require a
considerably larger amount of memory at \emph{computation} time.
For example, DAWA in two dimensions requires instantiating
requires instantiating a matrix representation of the workload \cite{dawa}.
For the workload that \overlook supports (the all-queries workload),
this requires a matrix of size $n^3$.
For such a workload, a relatively small domain with 1000 quantization intervals
would require at least 8 gigabytes of memory simply to compute the synopsis.

\subsection{Overlook performance}
\label{sec:scaling}

In this section, we benchmark the performance of \overlook
and demonstrate that adding differential privacy
does not substantially slow down the system or
change its underlying scaling properties.
In particular, the overall slowdown from privacy
is no greater than 2.5$\times$.

\subsubsection{Slowdown relative to public data}
\begin{figure}[t]
  \centering
  \begin{subfigure}{.49\columnwidth}
    \includegraphics[width=\columnwidth]{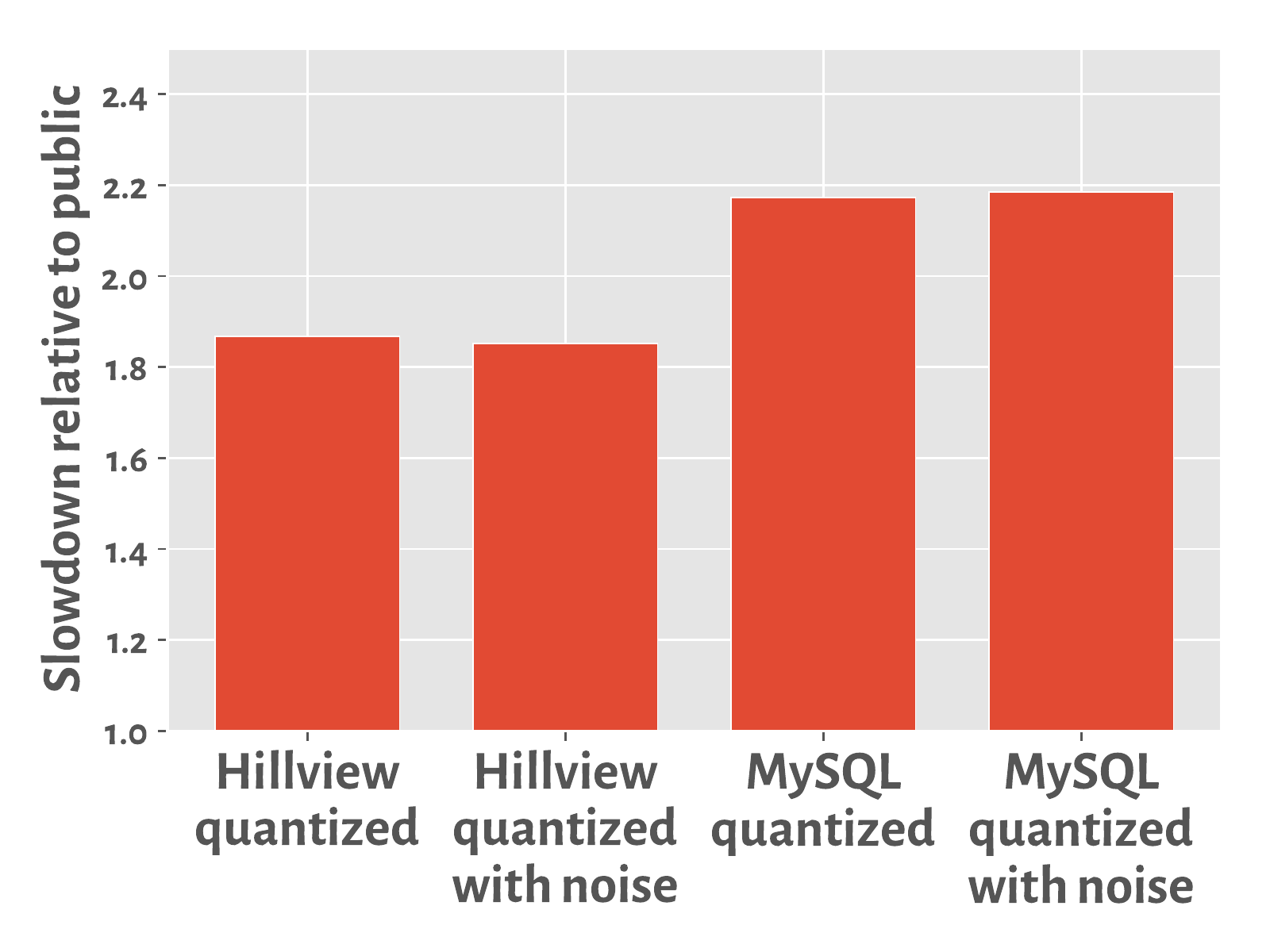}
    \caption{Histogram slowdown.}
  \end{subfigure}\hfill%
  \begin{subfigure}{.49\columnwidth}
    \includegraphics[width=\columnwidth]{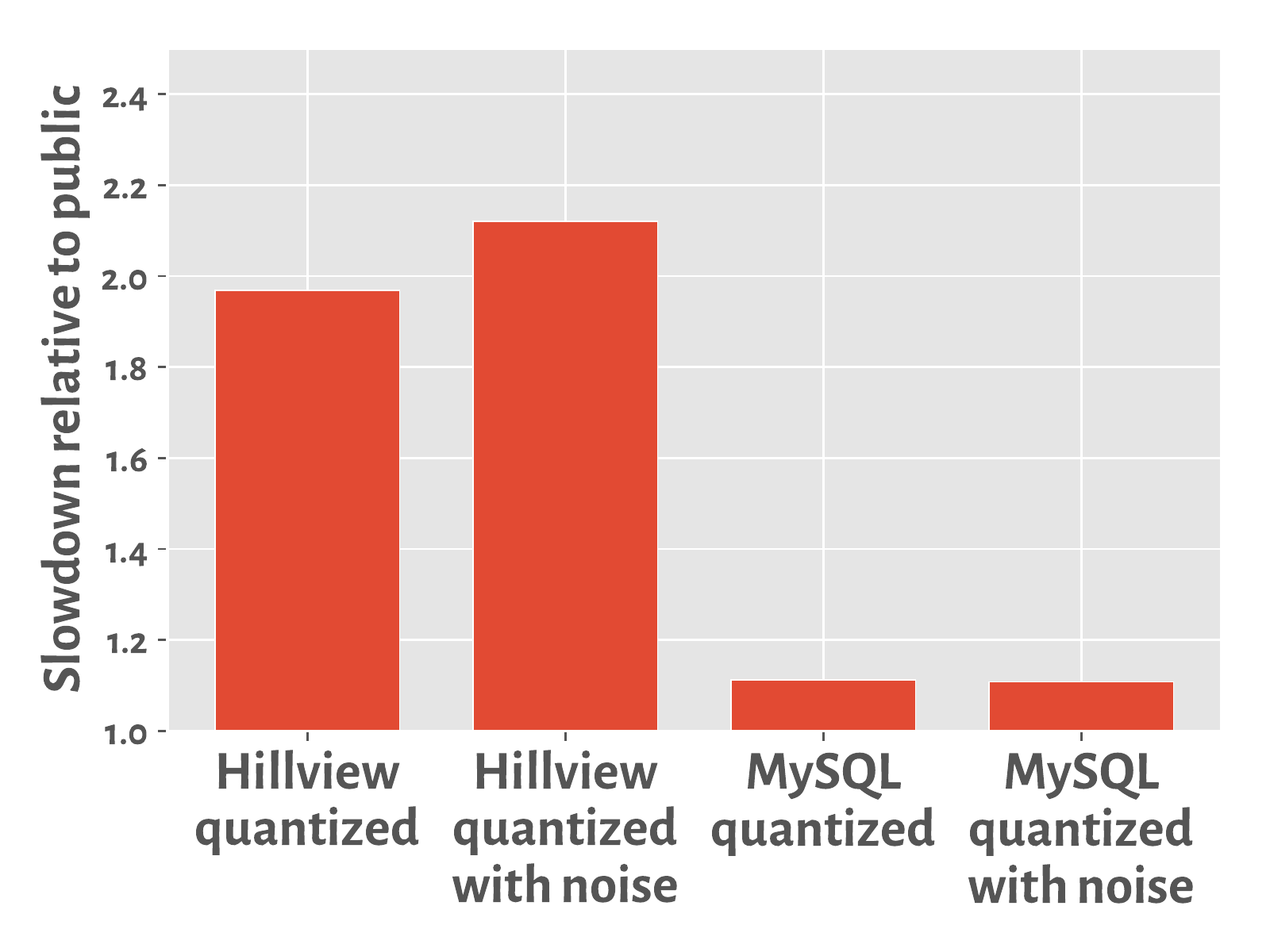}
    \caption{Heatmap slowdown.}
  \end{subfigure}
    \setlength{\belowcaptionskip}{-15pt}
  \caption{Slowdown relative to raw (non-private) databases
    for histograms and heat maps. In all cases, privacy adds at most a 2.5$\times$
    performance penalty.
    \label{fig:slowdown}}
\end{figure}

We first evaluate how much differential privacy
causes queries to slow down relative to queries on public data.
In order to understand the slowdown, we make two measurements for each backend:
first, the time required to quantize the dataset,
and second, the time required to answer a quantized histogram query with noise added.

Figure~\ref{fig:slowdown} shows the average slowdown when plotting histograms
and heat maps on the U.S. flights dataset using both the \hillview and MySQL backends.
The slowdown is below 2.5$\times$ for all configurations.
In all cases, the majority of the slowdown is a result of the quantization step.
This is intuitive: where each data point would initially have required one operation
to add it to the appropriate bucket,
quantization adds an additional operation to round the point to its nearest value in the public column domain.

\subsubsection{Scaling}
The \overlook frontend can be used with any SQL backend.
However, the \hillview distributed backend is powerful as it retains \hillview's
ability to scale to large datasets.

\begin{figure}[t]
  \centering
  \begin{subfigure}{.49\columnwidth}
    \includegraphics[width=\columnwidth]{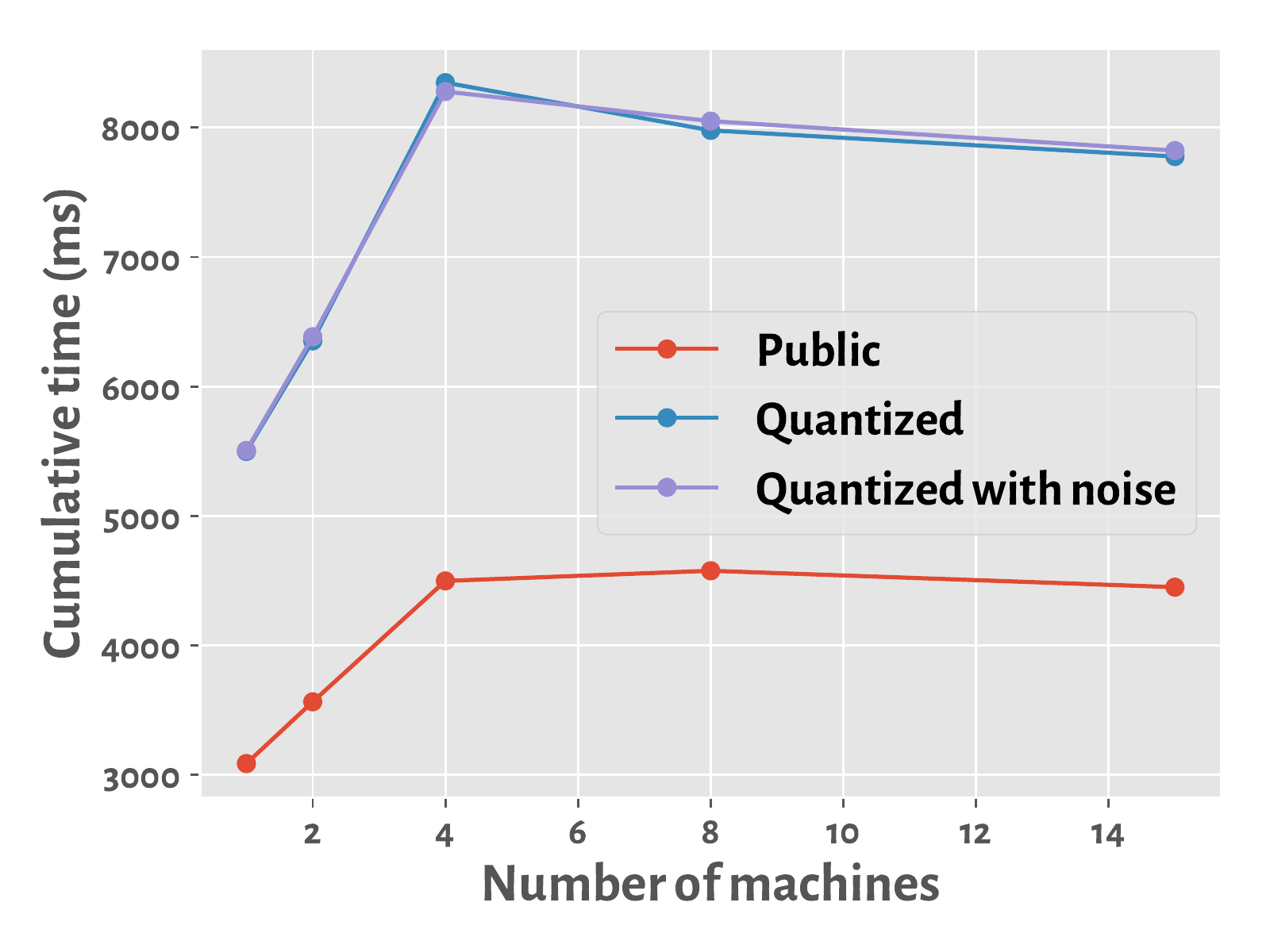}
    \caption{Histogram scaling.}
  \end{subfigure}\hfill%
  \begin{subfigure}{.49\columnwidth}
    \includegraphics[width=\columnwidth]{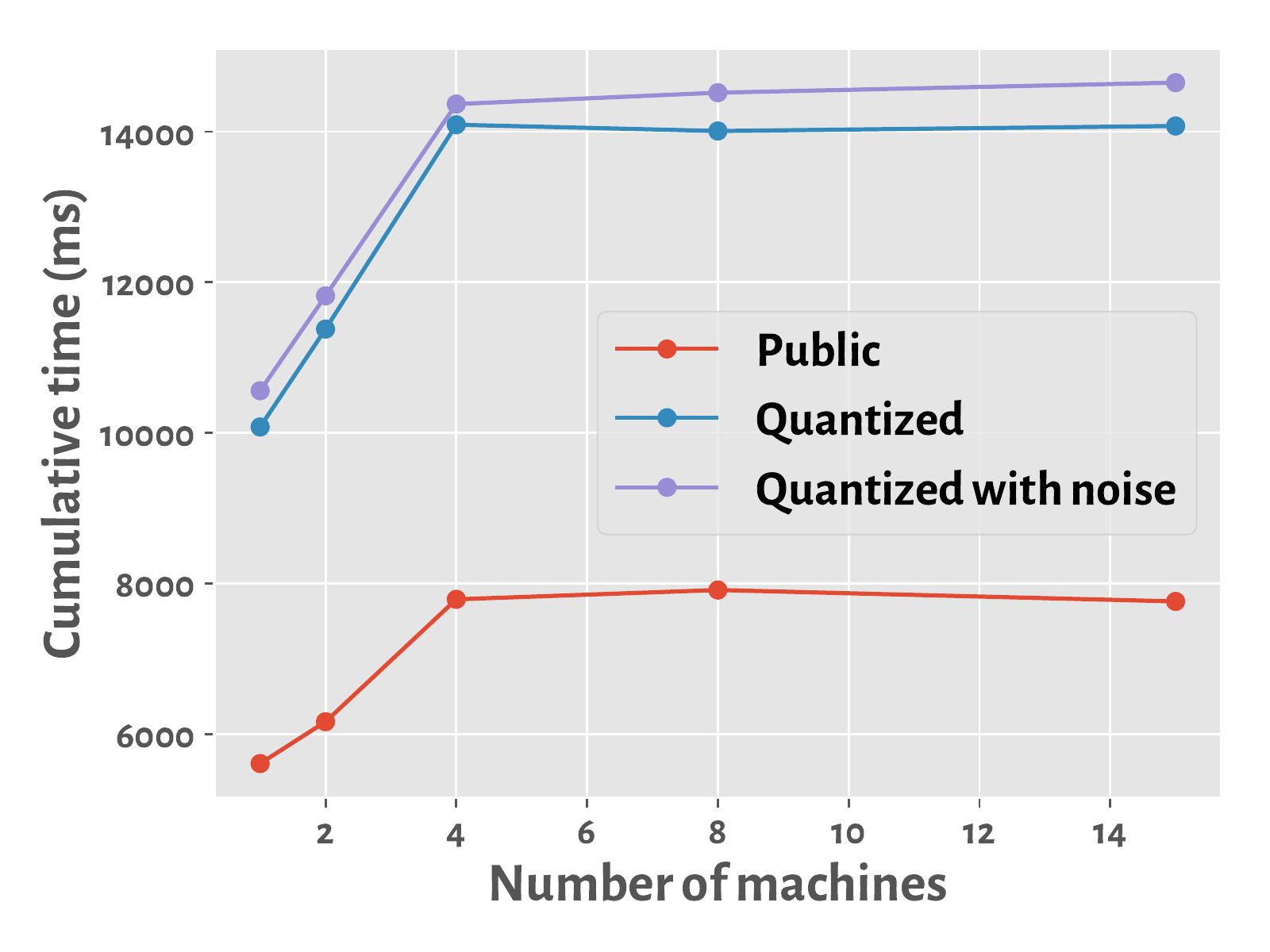}
    \caption{Heat map scaling.}
  \end{subfigure}
    \setlength{\belowcaptionskip}{-15pt}
  \caption{Average time to generate histograms for columns in the flights dataset
    as the number of machines grows. The data size grows with the number
    of machines, so the runtime remains constant.
    \label{fig:scaling}}
\end{figure}

We evaluate scaling using clusters of 1, 2, 4, 8,and 15 Amazon EC2 machines.
The total dataset size is 58.2 GB, split equally among the machines in the cluster.
so for linear scaling we expect the time required for each
query to be roughly the same regardless of the number of machines.
We measure time to compute charts once the data is already in memory.

In Figure~\ref{fig:scaling} we show our measurements that evaluate the
time breakdown for computing histograms over the U.S. flights dataset.
Each point corresponds to the total time required to compute a histogram or heat map for
every column or pair of columns.
The overhead of privacy is the same roughly 2$\times$ overhead as in Figure~\ref{fig:slowdown},
but privacy does not change the scaling behavior of the system at all, as expected.

\subsection{Visual error}
\label{sec:pixel-error}

At large enough data sizes, the error induced by differential privacy
can be \emph{smaller} than the pixel-level rounding error induced by the screen resolution.
In particular, in a 1-dimensional histogram,
\overlook rescales the $y$-axis to the maximum displayed value
in order to make use of all of the available vertical pixels.

Given $p$ vertical pixels and a maximum displayed value of $y$,
each pixel represents a count of $y/p$.
Hence, a confidence interval of size less than $y/p$ will be
smaller than a pixel on the screen.
Assuming the case of a single Laplace random variable added to each bucket
(i.e. the ``identity'' mechanism),
we ask what value of $\eps$ would suffice to achieve this level of error.

The inverse of the $\Lap(1/\eps)$ distribution is given by
\begin{align*}
  F^{-1}(x) = -(1/\eps)\sgn(p-0.5)\ln(1-2\vert x-0.5 \vert).
\end{align*}

Then we would like $F^{-1}(0.95) < y/p$ for a confidence level of $\alpha = 0.95$.
In this case, we arrive at an approximate solution of
$\eps > 2.303p/y$.
In other words, if the maximum count is approximately \textbf{2.3 times} the number of
vertical pixels,
a privacy level of $\eps=1$ will likely result in \emph{no visible difference} from the raw data.

\section{Related Work}

\paragraph{Differentially private database management systems}
A number of prior systems make differential privacy available through a SQL-like database API.
PINQ \cite{McSherry2009} implements a subset of SQL as well as a prototype visualization system
with incremental $\eps$-budgeting.
PINQ additionally pointed out that joins have potentially unbounded sensitivity,
and several later systems \cite{ProserpioGM14, restricted-sensitivity, djoin, flex}
propose methods for mitigating this issue.

Other work considers additional variants on SQL-like programming frameworks that allow
developers to easily express differentially private queries.
Airavat \cite{airavat} allows users to run custom MapReduce queries on sensitive data
by enforcing differential privacy on the queries.
Ektelo \cite{Zhang2018} exposes a number of higher-level operators as a programming framework
for differentially private mechanisms.
PrivateSQL \cite{privatesql} uses synopsis-based mechanisms rather than incremental budgeting
to release dataset views.
PrivateSQL introduces the notion of \emph{view sensitivity} as an approach to handle joins.
While \overlook does not explicitly target joins,
such techniques could naturally integrate with \overlook,
as the join is ultimately materialized as a tabular view of the data.
Chorus \cite{chorus} implements differential privacy directly in SQL.

Most of these prior systems support a broad set of queries written directly in SQL
or a SQL-like language.
In contrast, \overlook restricts queries to those visualizations 
enabled by the UI, which enables the release of a flexible and small synopsis.

\paragraph{Synopsis-based mechanisms}
A number of DP mechanisms \cite{salil-survey} are designed to support all queries in a given class of queries
simultaneously.
Recent work on synopses include the matrix mechanism \cite{li2010optimizing, li2012adaptive, li2015matrix},
wavelet transforms \cite{privelet},
and approaches that incorporate data-dependent partitioning \cite{xiao2012dpcube, xu2013differentially, qardaji2013differentially, cormode2012differentially}.
\overlook primarily relies on a hierarchical histogram \cite{Hay10, binary-mechanism}.
A growing body of work \cite{QardajiYL13, Qardaji:2014, dpbench} additionally considers data-dependent optimizations
that can improve the accuracy of synopses under certain query workloads.

A number of papers \cite{QardajiYL13, qardaji2013differentially, Hay10}
have investigated the accuracy of these methods in practice.
These papers support our claim that the synopses used in \overlook
give usable, and often optimal, accuracy in practice.

The idea of public quantization boundaries, or partitions, has been explored by PINQ \cite{pinq}
and FLEX \cite{flex}.
Both of these systems leave it to the data analyst,
rather than the data curator, to specify the quantization boundaries.

Similar ideas are used to analyze streaming data, \cf  \cite{Ghayyur0YMHMM18},  \cite{Chen2017}. A data aware version of the binary mechanism is presented in \cite{Acs:2012,dawa}.
It uses a private partitioning method that smooths regions of similar count.

\paragraph{Differentially private visualization}
PSI \cite{psi} may be the closest system to \overlook;
PSI makes $\eps$-budgeting more user-friendly by providing a visual interface for users
to interact with and understand the impact of various values of $\eps$.
In contrast to \overlook, PSI assumes a per-user, incremental $\eps$ budget;
additionally, PSI is not targeted toward the data exploration use case.

PINQ \cite{pinq} provides a case study of a differentially private map visualization as an application of the
framework.
\cite{chen2016differentially} provides methods to make linear and logistic regression plots differentially private.
VisDPT \cite{visdpt} is an interface to view two-dimensional trajectories in a differentially private manner.
\cite{zhangchallenges} studies the challenges involved in creating meaningful visualizations under differential privacy.

\section{Conclusion}

We have presented \overlook, a visualization system for private data that
provides interactive latencies both for data curators and data analysts.
\overlook's novel virtual synopsis enables it to scale to large
data domains while incurring minimal performance and storage overhead
over queries to raw data. \overlook can integrate with existing query
engines with no intrusive changes. \overlook makes differential privacy
accessible, useful, and performant, making it a practical privacy tool
for the real world.

\clearpage
\bibliographystyle{plain}
\bibliography{overlook}

\end{document}